\documentclass[aps,prx,amsmath,floats,floatfix,twocolumn,
  superscriptaddress,nofootinbib,showpacs]{revtex4-1}

\pdfoutput=1 
\usepackage{diagbox}
\usepackage{multirow}
\usepackage{braket}
\usepackage{amsfonts}
\usepackage{amsmath}
\usepackage{amssymb}
\usepackage{amsthm}
\usepackage{bm}
\usepackage{graphicx}
\usepackage{graphics}
\usepackage{latexsym}
\usepackage{rotating}
\usepackage[dvipsnames]{xcolor}
\usepackage{mathrsfs}
\usepackage{microtype}
\usepackage{verbatim}
\usepackage{url}

\usepackage[caption=false]{subfig}
\usepackage[normalem]{ulem}

\usepackage[breaklinks=true]{hyperref}
\hypersetup{
    colorlinks=true,
    linkcolor=NavyBlue,
    filecolor=Magenta,
    urlcolor=NavyBlue,
    citecolor=NavyBlue
}

\usepackage{yfonts}
\usepackage{float}
\usepackage{xspace} 
\usepackage{mathrsfs}
\usepackage{siunitx}
\usepackage{array}

\allowdisplaybreaks[4]

\newcommand{\be}{\begin{equation}}
\newcommand{\ee}{\end{equation}}

\newcommand{\ba}{\begin{align}}
\newcommand{\ea}{\end{align}}

\makeatletter
\newcommand*{\rom}[1]{\expandafter\@slowromancap\romannumeral #1@}
\makeatother

\AtBeginDocument{%
    \newwrite\bibnotes
    \def\bibnotesext{Notes.bib}
    \immediate\openout\bibnotes=\jobname\bibnotesext
    \immediate\write\bibnotes{@CONTROL{REVTEX41Control}}
    \immediate\write\bibnotes{@CONTROL{%
    apsrev41Control,author="08",editor="1",pages="1",title="0",year="1"}}
    \if@filesw
    \immediate\write\@auxout{\string\citation{apsrev41Control}}%
    \fi
}

\allowdisplaybreaks

\begin{document}


\title{Spectral method for metric perturbations of black holes: \\ Kerr background case in general relativity}

\author{Adrian Ka-Wai Chung}
\email{akwchung@illinois.edu}
\affiliation{Illinois Center for Advanced Studies of the Universe \& Department of Physics, University of Illinois at Urbana-Champaign, Urbana, Illinois 61801, USA}

\author{Pratik Wagle}
\email{pratik.wagle@aei.mpg.de}
\affiliation{Illinois Center for Advanced Studies of the Universe \& Department of Physics, University of Illinois at Urbana-Champaign, Urbana, Illinois 61801, USA}
\affiliation{Max Planck Institute for Gravitational Physics (Albert Einstein Institute), D-14476 Potsdam, Germany}

\author{Nicol\'as Yunes}
\affiliation{Illinois Center for Advanced Studies of the Universe \& Department of Physics, University of Illinois at Urbana-Champaign, Urbana, Illinois 61801, USA}

\date{\today}

\begin{abstract}
We present a novel approach, \textit{Metric pErTuRbations wIth speCtral methodS} (METRICS), to calculate the gravitational metric perturbations and the quasinormal-mode frequencies of rotating black holes of any spin without decoupling the linearized field equations. 
We demonstrate the method by applying it to perturbations of Kerr black holes of any spin, simultaneously solving all ten linearized Einstein equations in the Regge-Wheeler gauge through purely algebraic methods and computing the fundamental (corotating) quadrupole mode frequency at various spins.  
We moreover show that the METRICS approach is accurate and precise, yielding (i) quasinormal mode frequencies that agree with Leaver's, continuous-fraction solution with a relative fractional error smaller than $10^{-5}$ for all dimensionless spins below up to 0.95, and (ii) metric perturbations that lead to Teukolsky functions that also agree with Leaver's solution with mismatches below 1\% for all spins below 0.9.  
By not requiring the decoupling or the angular separation of the linearized field equations, the METRICS approach has the potential to be straightforwardly adapted for the computation of the quasinormal-mode frequencies of rotating black holes of any spin beyond general relativity or in the presence of matter. 
\end{abstract}

\maketitle


\section{Introduction}
\label{sec:intro}

Gravitational waves (GWs) are unique probes of fundamental physics because they are generated by the most energetic and extreme events in the Universe, such as the collision of black holes (BHs) and neutron stars \cite{LIGO_01, LIGO_02, LIGO_03, LIGO_04, LIGO_05, LIGO_06, LIGO_07, LIGO_08, LIGO_09, LIGO_10, LIGO_11, LIGO_GW190412, LIGO_GW190814}.
Unlike electromagnetic waves, GWs can propagate through most of the Universe unaffected by intervening matter, except when being lensed (see, e.g., \cite{Takahashi:2003ix}). 
GW detection, therefore, grants us unobscured access to physics in the highly-dynamical, and non-linear regime of gravity. In particular, this unobscured access to gravity can be used to test the validity of general relativity (GR) and to probe fundamental physics in extreme conditions \cite{LIGO_07, LIGO_11, LIGOScientific:2021sio, Perkins:2022fhr, Yunes:2013dva}. 

Even though Einstein's theory represents our best understanding of gravity and it has withstood all observational and experimental tests (see, e.g., ~\cite{Will2014,Stairs2003,Wex:2020ald,Yunes:2013dva,Will2014,Yagi:2016jml,Berti:2018cxi,Nair:2019iur,Berti:2018vdi}), GR still needs to be probed further.   
This is because over the past 50 years, theoretical and observational ``anomalies'' have somewhat challenged GR's validity.
On the theory front, GR predicts the existence of spacelike and timelike singularities, where the spacetime curvature diverges and GR's predicability is lost.
Such singularities are presumably resolved through quantum effects~\cite{Hawking:2015qqa, Almheiri:2019qdq, Giddings:1995gd, Preskill:1992tc, Kanitscheider:2007wq, Bena:2016ypk}, but, in spite of almost a century of theoretical efforts, Einstein's theory remains incompatible with quantum mechanics. 
Observationally, GR may require additional parity-violating physics (to satisfy the Sakharov conditions~\cite{Sakharov:1967dj}) to explain the matter-antimatter asymmetry of the early universe~\cite{Petraki:2013wwa, Gell-Mann:1991kdm, Alexander:2004us}, a fine-tuned cosmological constant~\cite{Nojiri:2006ri, Tsujikawa:2010zza} to explain the late-time acceleration of the Universe~\cite{late_time_acceleration_01, late_time_acceleration_02}, and dark matter to explain galaxy rotation curves~\cite{rotation_curve_01, rotation_curve_02}. 
These issues have prompted some to amend GR, inspiring a variety of modified gravity theories, such as dynamical Chern-Simon gravity~\cite{dCS_01, dCS_02, dCS_03} and Einstein-dilaton-Gauss-Bonnet gravity~\cite{EdGB_01, EdGB_02, EdGB_03, EdGB_04}, to name a few. 
To avoid repetition, we refer the reader to \cite{Chung:2023zdq} and references therein for a more comprehensive review of the motivation of modifying GR in the context of GW physics.

A better understanding of the aspects of GR that may require modification can be gained through GW tests of GR, which can yield new constraints on GR deviations, and maybe the detection of new anomalies. 
The approximately 100 GW signals detected by the advanced Laser Interferometric Gravitational-wave Observatory (LIGO) and Virgo detectors were mostly emitted by the coalescence of binary BHs~\cite{LIGO_01, LIGO_02, LIGO_03, LIGO_04, LIGO_05, LIGO_06, LIGO_07, LIGO_08, LIGO_09, LIGO_10, LIGO_11, LIGO_GW190412, LIGO_GW190814}. 
Of these events, $\sim 22$ contain clear ``ringdown'' signals, emitted after the BHs have collided and begun to settle down to their final stationary state through GW emission \cite{LIGOScientific:2021djp}. 
At late times, the ringdown can be described through quasinormal modes (QNMs), whose frequencies in GR are completely determined by the remnant's mass and spin (for astrophysical BHs). 
Modified gravity theories, however, also permit, in principle, BH solutions that are different from those in GR, and thus, may need additional parameters or ``hair'' (e.g.~functions of the coupling constants of the modified theory) to fully describe them (see e.g.~\cite{Yunes:2009hc,Yunes:2011we}).  
Moreover, the field equations in modified gravity are not Einstein's, and thus, the equations governing BH perturbations are also generically modified. 
These two facts lead to QNM spectra that can be very different from that of Kerr BHs in GR, and thus, their detection allows for new tests of Einstein's theory. 

Unfortunately, computing the gravitational QNM frequencies of rotating BHs in modified gravity is extremely challenging. 
This is because one needs to solve the non-Einsteinian field equations, linearized about a BH background that lacks the simplifying symmetries of the Kerr metric, typically leading to a very complicated set of coupled partial differential equations. 
Broadly speaking, the challenging task of computing QNM frequencies in modified gravity has so far been approached through two different perturbative techniques\footnote{In principle, one can also numerically simulate the ringdown phase of a BH formed by binary BH coalescence in modified gravity (see, e.g., \cite{AresteSalo:2023hcp}). 
However, numerical relativity in modified gravity is currently in its infancy, and more work is needed to ensure secular errors (due to the use of approximation methods) can be controlled in a gauge-independent way~\cite{Okounkova:2019dfo, Kovacs:2020ywu, Cayuso:2023aht}}.
The first approach works directly with metric perturbations, i.e.~linear deviations in the metric tensor with respect to the BH metric background. 
Using this ansatz in the field equations and linearizing about the BH background, one finds a complicated set of coupled, partial differential equations that one attempts to decouple into ``master equations'' for certain ``master functions'', such as those found by Regge and Wheeler \cite{Regge_Wheeler_gauge} and by Zerilli and Moncrief \cite{Zerilli_even, Moncrief:1974am} for perturbations of nonrotating BHs in GR. 
This metric perturbation approach has been successfully applied to nonrotating and to slowly rotating BHs in both GR and in modified gravity~\cite{Regge:PhysRev.108.1063,Zerilli:1971wd,Moncrief:1974am,Molina:2010fb,Cardoso:2009pk,Blazquez-Salcedo:2016enn,Wagle:2021tam,Pierini:2021jxd,Pierini:2022eim,Cano:2020cao,DeFelice:2023rra}. 
However, when studying perturbations of rotating BH spacetimes, the linearized field equations are too intricate, preventing the decoupling into master equations. 

The second approach works with curvature perturbations instead of metric ones,  relying on the Newman-Penrose (NP) formalism~\cite{Newman:1961qr}. 
In this approach, one first formulates the full Einstein equations and the Bianchi identities in terms of NP quantities, including spinor coefficients, Weyl scalars, and differential operators. 
With that at hand, one then linearizes these equations about the curvature of a BH background, using the symmetries of the latter (which forces certain spin coefficients to vanish) to simplify the resulting expressions. 
The simplified, linearized-curvature, Einstein and Bianchi equations can now be decoupled when the background is Petrov-type D (which is consistent with the symmetries of a Kerr BH metric), leading to the Teukolsky master equation~\cite{Teukolsky:1973ha} for a certain Teukolsky master function. 
The Teukolsky equation is a separable wave equation for the perturbed Weyl scalars $\Psi_0$ and $\Psi_4$, which represent ingoing and outgoing GW degrees of freedom~\cite{Teukolsky_01_PRL, Teukolsky:1973ha, Press:1973zz, Teukolsky:1974yv}.

This curvature perturbation approach has been (very recently) extended to incorporate modified gravity theories (with leading-order deviations from GR) in~\cite{Li:2022pcy,Hussain:2022ins}. 
This extension, the so-called ``modified Teukolsky formalism'', has been applied to rotating BHs in higher derivative gravity~\cite{Cano:2023jbk, Cano:2020cao, Cano:2021myl, Cano:2023tmv} and dynamical Chern-Simons gravity~\cite{Wagle:dcsslow}. 
Nonetheless, the modified Teukolsky formalism is complicated by the need to solve the Teukoslky equation twice: once in GR and a second time for the GR deviation of $\Psi_{0,4}$ with a very complicated source. 
The latter, in particular, requires metric perturbation reconstruction in GR~\cite{Li:2022pcy}, which can, in principle, be achieved through the use of the Hertz potential in the so-called Chrzanowski-Kogen-Kegeles (CKK) approach~\cite{Chrzanowski:1975wv, Toomani:2021jlo, Whiting_Price_2005, Lousto:2005xu, Ripley:2020xby, Ripley:2022ypi}. 
The need to solve a sourced Teukolsky equation, where the source depends on the reconstructed metric perturbation, can introduce numerical difficulties. 
Moreover, the perturbative extension of the Teukolsky approach assuming small deviations from GR can, in principle, be susceptible to secular errors that must be controlled. 

These difficulties motivate us to develop a method that can calculate the metric perturbations of a general BH, regardless of its spin and Petrov-type, based on spectral expansions. Henceforth, we shall refer to this as the \textit{Metric pErTuRbations wIth speCtral methodS} (METRICS) approach. 
Given the tremendous difficulty of this problem in modified gravity, a sensible first step is to map the path forward by developing this approach within Einstein's general theory. 
Building on other related work on spectral methods~\cite{Jansen:2017oag, Eperon:2019viw, Dias:2018ufh, Ripley:2022ypi, Cardoso:2013pza, Ripley:2020xby, Dias:2015wqa, Dias:2021yju, Dias:2022oqm, Jansen:2017oag, Langlois:2021aji, Langlois:2021xzq, Dias:2014eua, Dias:2014eua, Spectral_03, Spectral_04, Spectral_05, Santos:2015iua}, we did precisely this in \cite{Chung:2023zdq} by focusing on perturbations of a nonrotating (Schwarzschild) BH in GR.
The goal of this paper is to generalize \cite{Chung:2023zdq} to perturbations of rotating (Kerr) BHs in GR, allowing the BH background to have arbitrary spin. As we will see, the generalization to BHs with arbitrary spins in GR will require us to improve the METRICS approach of~\cite{Chung:2023zdq} in various ways, which will be crucial when, in the future, this approach is applied to BHs perturbations in modified gravity. 

We extend the METRICS approach to Kerr BHs in GR as follows.
We begin by deriving the linearized Einstein field equations that govern the metric perturbations of a Kerr BH in the Regge-Wheeler gauge (Sec.~\ref{sec:EFEs}), a commonly adopted gauge to study gravitational BH perturbations. 
We then construct an ansatz for the metric perturbations that asymptotes to the desired boundary conditions at spatial infinity (outgoing) and at the horizon (ingoing) in outgoing and ingoing Kerr null coordinates (Sec.~\ref{sec:ansatz}).
This ansatz, of course, does not solve the linearized Einstein equations, so we multiply it with unknown, finite functions, which we then spectrally expand into products of Chebyshev polynomials (of a compactified Boyer-Lindquist radial coordinate) and associated Legendre polynomials (of the Boyer-Lindquist polar angular coordinate).
The spectral product decomposition transforms the linearized Einstein field equations into a system of homogeneous, linear and algebraic equations (Sec.~\ref{sec:Conversion}). 
The quadratic eigenvalues of the linear algebraic equations correspond to the QNM frequencies of the Kerr BH.
The eigenvector of the QNM frequencies can be used to reconstruct metric perturbations swiftly. 

The general extension described above is the obvious way to extend the METRICS approach from a Schwarzschild to a Kerr background, but further refinements are needed to obtain sufficiently accurate and precise QNM frequencies. 
The first refinement is intended to achieve numerically stability in the QNM frequency calculation. 
In the nonrotating case, we converted six (out of the ten) linearized Einstein equations into a quadratic eigenvalue problem via spectral expansions, and then, we transformed the quadratic eigenvalue problem into a linear generalized eigenvalue problem. 
For a Kerr BH background, however, this is insufficient to obtain accurate QNM frequencies because the solution we converge to need not satisfy the remaining four linearized equations. 
Transforming the quadratic eigenvalue problem into a linear generalized eigenvalue problem also introduces significant numerical instability, which undermines the accuracy of the QNM frequency calculation. 
We therefore refine the METRICS approach by developing a Newton-Raphson algorithm to simultaneously solve all the linear, homogenous algebraic equations that result from the ten, linearized Einstein equations (Sec.~\ref{sec:newton_method}). 

To initiate the Newton-Raphson iterations, we need an initial guess, which we here choose as the first two significant digits of the QNM frequencies of a Kerr BH, and zero for the remaining, initial spectral coefficients\footnote{In modified gravity, this choice of initial guess would not be possible since the QNM frequencies is precisely what one wishes to calculate in the first place. 
However, for modifications to GR that are small deformations, one can still initialize the Newton-Raphson algorithm at the GR values of the QNM frequencies, and then allow the algorithm to explore deviations, a generalization we leave to future work.}; we also explored other, more general choices for the initial guess of the QNM frequencies to find that the Newton-Raphson method still finds the correct solution, although it takes more iterations. 
The Newton-Raphson method requires the inversion of a coefficient matrix that is rectangular (not square), because the linearized Einstein equations are ten, while the metric perturbations are characterized by six functions in the Regge-Wheeler gauge. 
The rectangular coefficient matrix cannot be inverted as done usually for a square matrix, a problem we circumvent through the use of the Moore-Penrose inverse, a generalization of the square-matrix inverse to a rectangular matrix.

We apply this generalized and refined METRICS approach to accurately compute the QNM frequencies and subdominant modes of Kerr BHs with dimensionless spins up to $0.95$, as we will show in Sec.~\ref{sec:results}. 
In particular, we show that we can calculate the QNM frequencies of the fundamental corotating quadrupole mode (the so-called ``022" mode) of the Kerr BH of dimensionless spin $\leq 0.95 $ with a relative fractional error in the real and imaginary parts that is less than $10^{-5}$.
This relative fractional error is significantly smaller than the current measurement uncertainty of the relative fractional departure of the 022-mode frequency from its GR prediction (of $\mathcal{O}(10^{-2})$), obtained by combining all the astrophysical ringdown signals detected by the advanced LIGO and Virgo detectors~\cite{LIGOScientific:2021sio}, as well as the projected measurement uncertainty obtained by combining the detections made by next-generation detectors (of $\mathcal{O}(10^{-4})$) \cite{Maselli:2023khq}.
Thus, the QNM frequency computed using the METRICS approach is both precise and accurate enough to be applied in the analysis of ringdown signals detected by existing and future ground-based detectors. 

Another important benefit of the METRICS approach is its ability to solve for the metric perturbations directly. 
This allows us to validate the METRICS approach by using these metric perturbations to compute the perturbed Weyl scalars $\Psi_{0}$ and $\Psi_{4}$, and then verify that they satisfy the Teukolsky equation. 
Indeed, in Sec.~\ref{sec:metric_numerical_validation}, we find that the perturbed Weyl scalar obtained from the metric perturbations solved with the METRICS approach has a ``mismatch'' with respect to that obtained from Leaver's, continuous fraction method on the equatorial plan that is below $\sim 10^{-2}$ .
Leaver's continuous fraction perturbed Weyl scalar also allows us to validate our metric ansatz by reconstructing the metric perturbation through the CKK approach. 
We implement this reconstruction and find that, overall, our ansatz is consistent with the metric reconstruction. 
Through these validations, we ensure that the generalization and refinement of the METRICS approach developed in this paper are ready for deployment in modified gravity theories, a task that is left to future work.  

The computational resources required to apply the METRICS approach to accurately compute the metric perturbations of a Kerr black hole are also reasonable. 
Using a symbolic computational package, such as \texttt{Mathematica}, and a single central processing unit on a standard laptop computer, the Newton-Raphson algorithm takes about $\sim$ 1200 seconds (approximately 20 minutes) to compute the 022-mode frequency with 30 Chebyshev and associated Legendre polynomials. 
The computational time can be significantly reduced if one optimizes the calculation, using, for example, other computational languages (such as C/C++), parallel computations, or graphical processing units for execution. 
We estimate that these enhancements can easily accelerate the METRICS calculations to under a second, but we leave such optimizations for future explorations. 

The remainder of this paper deals with the computational details that lead to the results described above and it is organized as follows.  
In Sec.~\ref{sec:EFEs}, we review the Kerr metric, the Regge-Wheeler gauge conditions, the asymptotic behavior of metric perturbations at the event horizon and future null infinity, and we spectrally expand the metric perturbations. 
In Sec.~\ref{sec:line-EEs}, we discuss the mathematical structure of the linearized Einstein equations and convert them into a system of linear homogenous algebraic equations. 
In Sec.~\ref{sec:extraction}, we describe the details of the refined procedures to extend the METRICS approach to the Kerr BH, including the Newton-Raphson algorithm. 
In Sec.~\ref{sec:results}, we calculated the QNM frequencies of the Kerr BH extracted with the METRICS approach.
The robustness of the QNM frequencies obtained using the METRICS approach is studied in Sec.~\ref{sec:robustness}. 
We show that METRICS can correctly reconstruct metric perturbations around the Kerr BH in Sec.~\ref{sec:metric_reconstructions}. 
Finally, we conclude the paper in Sec.~\ref{sec:conclusion} by exploring possible future applications of the METRICS approach to study the gravitational QNMs of different BHs and theories of gravity.  

Henceforth, we assume the following conventions: 
$x^{\mu} = (x^0, x^1, x^2, x^3) = (t, r, \chi, \phi)$, where $\chi = \cos \theta$ and $\theta$ is the polar angle;
the signature of the metric tensor is $(-, +, +, +)$;
gravitational QNMs are labelled by $n l m$ or $(n, l, m)$, where $n$ is the principal mode number, $l$ is the azimuthal mode number \footnote{Note that $l$ is, in general, different from $\ell$, the degree of the associated Legendre polynomials in the product decomposition of the metric perturbation functions. Although these numbers are the same for a Schwarzschild BH background, this is not so for a Kerr BH background.} and $m$ is the magnetic mode number of the QNM;
the QNM frequencies computed using the METRICS approach are referred to as ``METRICS QNM frequencies"; 
Greek letters in index lists stand for spacetime coordinates;
for the convenience of the reader, we have presented a list of all definitions and symbols in Appendix~\ref{sec:Appendix_A}.

\section{Metric perturbations in a Kerr BH background}
\label{sec:EFEs}

In this section, we discuss our representation of the background Kerr spacetime, derive the asymptotic behavior of the metric perturbations at the event horizon and spatial infinity, and then conclude with a quick description of the spectral expansion of the metric perturbations.

\subsection{Background spacetime}
\label{sec:metpert}

The solution to the vacuum Einstein equation $G_{\mu \nu} = 0$ that represents a stationary, rotating and uncharged BH is the Kerr metric denoted by $g_{\mu \nu}^{(0)}$. The line element associated with this metric can be written in Boyer-Lindquist coordinate
as \cite{Teukolsky:2014vca}, 
\begin{equation}\label{eq:metric}
\begin{split}
ds^2 &= g_{\mu \nu}^{(0)} dx^{\mu}  dx^{\nu} \\
& = - \left( 1-\frac{2 M r}{\Sigma} \right) dt^2 - \frac{4 M^2 a r}{\Sigma} (1 - \chi^2) d \phi dt \\
& \quad + \frac{\Sigma}{\Delta} dr^2 + \frac{\Sigma}{1 - \chi^2} d \chi^2 \\
& \quad + \left[r^{2} + M^2 a^{2}+\frac{2 M^3 a^{2} r}{\Sigma} (1 - \chi^2)\right](1-\chi^2) d\phi^2, \\
\end{split}
\end{equation}
where recall that $\chi = \cos{\theta}$ and we have defined
\begin{align} \label{eq:metric_quantities}
\Sigma &= r^2 + M^2 a^2 \chi^2, \nonumber\\
\Delta &= (r-r_+)(r-r_-), \nonumber\\
r_{\pm} &= M(1 \pm b), \nonumber\\
b & = \sqrt{1-a^2},
\end{align}
where $r_{\pm}$ are the locations of the inner and outer event horizons, $M$ is the Kerr BH mass, which is taken to be one throughout this paper, and $0\leq a<1$ is the dimensionless spin.
Let us also define here two variables associated with the Kerr spacetime that will be frequently used in the subsequent calculations. 
The first is the angular velocity of the outer event horizon, 
\begin{equation}
\Omega_H = \frac{a}{2r_+}. 
\end{equation}
The second is the tortoise coordinate of the Kerr BH, defined as \cite{Teukolsky_01_PRL, Teukolsky:1973ha, Teukolsky:1974yv}
\begin{equation}\label{eq:r_*}
\frac{d r_*}{d r} = \frac{r^2 + M^2 a^2}{\Delta}. 
\end{equation} 
Explicit integration then yields 
\begin{equation}
\begin{split}
r_* = r + M\Bigg[ & \frac{1+b}{b}\log\left( \frac{r-r_+}{2M}\right) \\
& - \frac{1-b}{b}\log\left( \frac{r-r_-}{2M}\right) \Bigg]. 
\end{split}
\end{equation}

\subsection{Metric decomposition and linearized Einstein equations}

We now consider linear perturbations of the metric tensor, such that
\be \label{eq:metpert}
g_{\mu\nu} = g_{\mu\nu}^{(0)} + \epsilon \; h_{\mu\nu} \,,
\ee
where $g_{\mu\nu}^{(0)}$ is the background metric of Eq.~\eqref{eq:metric}, $h_{\mu\nu}$ is the metric perturbations, and $\epsilon$ is a bookkeeping parameter for the perturbations. 
The metric perturbations $h_{\mu\nu}$ can be decomposed into axial and polar parts~\cite{Regge:PhysRev.108.1063, Berti_02, Zerilli:1971wd, Moncrief:1974am},
\be
\begin{split}
h_{\mu\nu} (t,r,\chi,\phi) = & h^{\rm odd}_{\mu\nu} (t,r,\chi,\phi) + h^{\rm even}_{\mu\nu} (t,r,\chi,\phi) \,,
\end{split}
\ee
where
\begin{widetext}
\begin{subequations} \label{eq:metpert}
\be \label{eq:odd}
	 h^{\rm odd}_{\mu\nu} (t,r,\chi,\phi) = e^{im\phi-i\omega t}
\begin{pmatrix}
    0 & 0 & -im(1-\chi^2)^{-1} h_5(r,\chi) &  (1-\chi^2) \partial_\chi h_5(r,\chi) \\
    * & 0 & -im(1-\chi^2)^{-1} h_6(r,\chi) &  (1-\chi^2) \partial_\chi h_6(r,\chi) \\
	* & * & 0 & 0  \\
	* & * & * & 0
\end{pmatrix}
\,,
\ee
and
\be \label{eq:even}
h^{\rm even}_{\mu\nu} (t,r,\chi,\phi) = - e^{im\phi-i\omega t}
\begin{pmatrix}
	h_1(r,\chi) & h_2(r,\chi) & 0 &  0 \\
	* & h_3(r,\chi) & 0 &  0 \\
	* & * & \left( 1 - \chi^2 \right)^{-1} h_4(r,\chi) & 0  \\
	* & * & *  & \left( 1 - \chi^2 \right) h_4(r,\chi)
\end{pmatrix}
\,,
\ee
\end{subequations}
\end{widetext}
and where we have made use of the Regge-Wheeler gauge~\cite{Regge:PhysRev.108.1063, Berti_02} for the Kerr BH.  The quantities $h_{k}$ with $k \in (1,6)$ are, so far, unknown functions of $r$ and $\chi$. Note that we have assumed that the parity sectors of the metric perturbations that are purely ingoing at the horizon and outgoing at future null infinity depend on the same QNM frequency, because of isospectrality in GR. This assumption would obviously have to be relaxed if working in modified gravity, where isospectrality breaks~\cite{Wagle:2021tam,Pierini:2021jxd,Molina:2010fb,Cardoso:2009pk,Blazquez-Salcedo:2016enn,isospectrality-paper}. Note also that we have chosen a specific ansatz for the polar and axial metric perturbations, guided by the Regge-Wheeler gauge and the angular structure of the ansatz for metric perturbations around slowly rotating BHs. 

\subsection{Asymptotic behavior of the metric perturbations}
\label{sec:ansatz}

According to the methodology presented in \cite{Chung:2023zdq}, we should first substitute Eq.~\eqref{eq:metpert} into the Einstein equations and linearize in $\epsilon$ to obtain the linearized Einstein equations for $h_{\mu \nu}$. We would then solve the linearized Einstein equations asymptotically at spatial infinity and at the event horizon (i.e.,~at the bifurcation two-sphere).
Unfortunately, for the Kerr BH background case, the resulting linearized field equations are too complicated to be diagonalized using the methods presented in~\cite{Langlois:2021xzq}, because the Kerr spacetime is just too complicated.
Instead, we proceed by studying the null coordinates defined by the principal null geodesics of the background spacetime, and the virtue of the METRICS approach that the computed QNM frequencies do not sensitively depend on the choice of radial scaling we choose in the ansatz for the radial function (i.e. the accuracy of the METRICS approach does not sensitively depend on the chosen $ \rho_H^{(k)}$ and $\rho_{\infty}^{(k)}$, which will, respectively, be first defined by Eqs.~\eqref{eq:asymptotic_limits1} and \eqref{eq:asymptotic_limits2}, see also Sec. VI. B. of \cite{Chung:2023zdq}).

We first study the asymptotic behavior at the event horizon, where metric perturbations are purely ingoing, and thus, their wavefronts should follow the principal null geodesics that are ingoing at the horizon \cite{Dias:2010eu, Dias:2014eua}.
The geodesics are more suitably described in ingoing Kerr null coordinates $(v, r, \theta, \varphi)$, where \cite{Poisson:2009pwt}
\begin{equation}
\begin{split}
v & = t + r_*, \\
\varphi & = \phi + \tilde{r}, 
\end{split}
\end{equation}
and where $\tilde{r}$ is defined by 
\begin{equation}
\frac{d \tilde{r}}{dr} = \frac{M^2 a}{\Delta} \Rightarrow \tilde{r} = \frac{M a}{2b} \log \left( \frac{r-r_+}{r-r_-}\right). 
\end{equation}
As metric perturbations should be regular in these coordinates, the asymptotic behavior of $h_k$ near the event horizon, for a fixed $\chi$, must be 
\begin{equation}
\label{eq:asymptotic_limits1}
\begin{split}
\lim \limits_{r \rightarrow r_+} h_k (r, \chi) & \propto \lim \limits_{r \rightarrow r_+}  e^{- i \omega v + i m \varphi}\\
& \sim \left( r-r_+ \right)^{-i (\omega - m \Omega_{H}) \frac{1+b}{b} - \rho_H^{(k)}} \\
& \quad \times \sum_{p=0}^{\infty} b_{p} (r-r_+)^{p}, 
\end{split}
\end{equation}
where $b_{p}$ are constants and $\rho_{H}^{(k)}$ is an $k$-dependent parameter controlling the divergent behavior of $h_{k}$ at $r=r_+$ \cite{Chung:2023zdq}.

We now study the asymptotic behavior at spatial infinity, where metric perturbations should be purely outgoing, and thus, their wavefronts should follow the principal null geodesics that are outgoing. 
These geodesics are more suitably described by outgoing Kerr null coordinates $(u, r, \theta, \varphi)$, where
\begin{equation}
\begin{split}
u & = t - r_*, \\
\varphi & = \phi - \tilde{r}. 
\end{split}
\end{equation}
As metric perturbations should also be regular in these coordinates, the asymptotic behavior of $h_k$ near spatial infinity, at a fixed $\chi$, is
\begin{equation}
\label{eq:asymptotic_limits2}
\begin{split}
\lim \limits_{r \rightarrow + \infty} h_k (r, \chi)  & \propto \lim \limits_{r \rightarrow + \infty} e^{- i \omega u + i m \varphi}\\
& \sim e^{i \omega r} r^{2 i M \omega  + \rho_{\infty}^{(k)}} \sum^{\infty}_{p=0}\frac{a_{p}}{r^p},  
\end{split}
\end{equation}
where $a_{p}$ are constants and $\rho_{\infty}^{(k)}$ is another $k$-dependent parameter controlling the divergent behavior of $h_{k}$ at spatial infinity.
The above is also the asymptotic behavior at spatial infinity of metric perturbations around the Schwarzschild BH \cite{Chung:2023zdq}.
This is reasonable because, at spatial infinity, the background spacetime of both the Schwarzschild and Kerr BHs reduces to the Minkowski spacetime.

Although the arguments presented above are only valid for vacuum rotating BHs in GR, we expect a similar argument to hold in more general situations. In modified gravity, the asymptotic behavior of the metric perturbations cannot necessarily be obtained by studying the first-order form of the (null) geodesic equations. Such a first order form may not exist if the background spacetime is of Petrov type I and does not possess a Killing tensor or a Carter-like constant. Instead, one would have to determine the principal null directions of the spacetime by looking at the independent roots of the Weyl tensor contracted onto a certain antisymmetric, exterior product of four copies of the null tangent vector, which defines the Petrov class, as done e.g.~in~\cite{Owen:2021eez} for dynamical Chern-Simons gravity. 

\subsection{Ansatz for the metric perturbations}
\label{sec:ansatz}

The results above motivate us to resum and peel off the asymptotic behaviors of $h_k(r, \chi)$ through the following product decomposition
\be \label{eq:radspec}
h_k(r, \chi) = A_k(r)  u_k(r, \chi) \,,
\ee
where $u_k(r, \chi)$ are correction functions that are finite for all $r\in[r_+, +\infty]$. 
In Eq.~\eqref{eq:radspec}, $A_k(r)$ is the ``asymptotic controlling factor'' of $h_k(r, \chi)$, which we define as
\begin{equation}\label{eq:asym_prefactor}
A_k(r) = e^{i \omega r} r^{2 i M \omega + \rho_{\infty}^{(k)}} \left( \frac{r-r_+}{r}\right)^{-i M (\omega - m \Omega_{H}) \frac{1+b}{b}- \rho_H^{(k)}}\,,
\end{equation}
This function has the property that it approaches Eq.~\eqref{eq:asymptotic_limits1} near the horizon, while it approaches Eq.~\eqref{eq:asymptotic_limits2} near spatial infinity. 
In Eq.~\eqref{eq:radspec}, $u_k(r, \chi)$ is then a correction factor that is not only bounded, but also has trivial boundary conditions, approaching a finite function of $\chi$ both at the event horizon and at spatial infinity.  

The parameters $\rho_{H}^{(k)}$ and $\rho_{\infty}^{(k)}$ in Eq.~\eqref{eq:asym_prefactor} control the divergence of the metric function at the event horizon and at spatial infinity, and thus, these constants should depend on the index $k$ and the dimensionless spin $a$. 
For the time being, we will assume that $\rho_{H}^{(k)}$ and $\rho_{\infty}^{(k)}$ are the same $\rho_{H}^{(k)}$ and $\rho_{\infty}^{(k)}$ of the Schwarzschild BH\footnote{Note that these $\rho_{H}^{(k)}$ and $\rho_{\infty}^{(k)}$ are different from those given in \cite{Chung:2023zdq} because in the latter we chose a different normalization convention for $h_{1, 2, 3, 4}$ (see Eq. (5b) of \cite{Chung:2023zdq}).}, 
\begin{equation}\label{eq:rhos}
\begin{split}
\rho_H^{(k)} & = 
\begin{cases}
2, ~~ \text{for $k = 3$ }\\
1, ~~ \text{for $k = 2\text{  or  } 6$ }\\
0, ~~ \text{otherwise}
\end{cases}, \\
\rho_{\infty}^{(k)} & = 
\begin{cases}
2, ~~ \text{for $k \neq 4$ }\\
1, ~~ \text{for $k = 4$}
\end{cases}.
\end{split} 
\end{equation}
The numerical results presented in Sec.~\ref{sec:results} show that the Schwarzschild values for $\rho_{H}^{(k)}$ and $\rho_{\infty}^{(k)}$ indeed allow for accurate and robust computations of the Kerr QNM frequencies. 
In Sec.~\ref{sec:robustness}, we will study different choices of $\rho_{H}^{(k)}$ and $\rho_{\infty}^{(k)}$ to address the possibility that $\rho_{H}^{(k)}$ and $\rho_{\infty}^{(k)}$ vary with $a$. 

Since the correction function $u_k(r, \chi)$ is bounded, we can spectrally expand it, but before doing so, we must transform its independent variable to a compactified coordinate.
Following \cite{Chung:2023zdq}, we define a compactified radial coordinate, $z$, via \cite{Langlois:2021xzq, Jansen:2017oag}
\begin{equation}\label{eq:z}
z(r) = \frac{2r_+}{r} - 1, 
\end{equation}
so that $u_k$ is a bounded function in the finite domain $z \in [-1, +1] $. With this in hand, the correction factor can now be spectrally expanded into a basis function, which we here choose to be the product of Chebyshev and associated Legendre polynomials, such that 
\begin{equation}
\label{eq:spectral_decoposition_correction_factor}
u_k (z, \chi) = \sum_{n=0}^{\infty} \sum_{\ell=|m|}^{\infty} v_k^{n \ell}  T_{n}(z) P^{|m|}_{\ell}(\chi)\,,
\end{equation}
where $ v_k^{n \ell}$ are constant coefficients.
As in \cite{Chung:2023zdq}, we could have chosen a different basis function for the spectral expansion, as long as it is complete and orthogonal. 
As we will see, the product of Chebyshev and associated Legendre polynomials is sufficient for our purposes, and other choices can be studied elsewhere.  

We can now put all of these results together to write a final expression for the spectrally decomposed metric perturbation
$h_k(z, \chi)$, namely
\begin{equation}\label{eq:spectral_decoposition_factorized}
h_k (z, \chi) = A_k(z) \sum_{n=0}^{\infty} \sum_{\ell=|m|}^{\infty} v_k^{n \ell}  T_{n}(z) P^{|m|}_{\ell}(\chi). 
\end{equation}
Although Eq.~\eqref{eq:spectral_decoposition_factorized} gives the full spectral expansion of the metric perturbation along the angular coordinate $\chi$ and the compactified spatial coordinate $z$, it is in practice impossible to include an infinite number of terms in the sums.
Instead, we must truncate the expansion at some $\mathcal{N}_z$ for the Chebyshev sum and some $\mathcal{N}_{\chi}$ for the associated Legendre sum, 
\begin{equation}\label{eq:spectral_decoposition_factorized_finite}
h_k (r, \chi) = A_k(r) \sum_{n=0}^{\mathcal{N}_z} \sum_{\ell=|m|}^{\mathcal{N}_{\chi}+|m|} v_k^{n \ell}  T_{n}\left[z(r)\right] P^{|m|}_{\ell}(\chi)\,,
\end{equation}
In what follows, we will use this truncated spectral expansion to compute the QNM frequencies of the Kerr BH. We will keep $\mathcal{N}_z$ and $\mathcal{N}_{\chi}$ independent for now, but when evaluating QNM frequencies, we will go ``along the diagonal'' of this sum by setting $\mathcal{N}_z = \mathcal{N}_{\chi} = \mathcal{N}$.

\section{Linearized Einstein equations about a Kerr BH background as an algebraic problem in the METRICS approach}
\label{sec:line-EEs}

In this section, we will derive the linearized Einstein equations that must be satisfied by the finite correction functions $u_{k}$, which we will later reduce to an algebraic system of equations for the $v_{k}^{n \ell}$ coefficients and the QNM frequencies. 

\subsection{Linearized Einstein equations}

With the decomposition (Eq.~\eqref{eq:metpert}) defined, we can now find the system of equations that the metric perturbations $h_k(r,\chi)$ must satisfy. 
Following \cite{Chung:2023zdq}, we do not treat the odd and even perturbations separately.

We derive the linearized Einstein equations of $h_k(r, \chi) $ by substituting Eq.~\eqref{eq:metpert} (in the form of Eq.~\eqref{eq:spectral_decoposition_factorized_finite}) into the trace-reversed Einstein equation 
\begin{equation}\label{eq:EFEs}
R_{\mu}{}^{\nu} = 8 \pi \left( T_{\mu}{}^{\nu} - \frac{1}{2} \delta_{\mu}{}^{\nu} T \right), 
\end{equation}
where $R_{\mu}{}^{\nu} = g^{\nu \alpha} R_{\mu \alpha}$, and $T = g^{\alpha \beta} T_{\alpha \beta}$ is the trace of the energy-momentum tensor. 
We linearize this form of the Einstein equation because the resulting equations are significantly shorter than those derived from $G_{\mu \nu} = 8 \pi T_{\mu \nu}$. 
In this paper, we are concerned only with vacuum perturbations, and thus $T_{\mu \nu} = 0$. 
Linearizing Eq.~\eqref{eq:EFEs}, one finds a system of ten coupled, partial differential equations for the six unknown functions $h_k(r,\chi)$. 

We now massage the linearized Einstein equations to cast them in a form that is more amenable to a spectral expansion. 
We note that the components of the background metric tensor $g_{\mu\nu}^{(0)}$ in Boyer-Lindquist coordinates, whose line element is in Eq.~\eqref{eq:metric}, are rational functions of $r$ and $\chi$. 
Therefore, the coefficient functions multiplying the metric perturbations $h_k$ in the linearized Einstein equations must also be rational functions of $r$ and $\chi$, since they can only depend on background quantities and their derivatives. 
With this understanding, we can always express the $k$-th linearized field equation,
after appropriate factorization and multiplication through common denominators, as
\begin{equation}\label{eq:pertFE-1}
\begin{split}
& \sum_{j=1}^{6} \sum_{\alpha, \beta = 0}^{\alpha + \beta \leq 3} \sum_{\gamma=0}^{2} \sum_{\delta=0}^{d_{r}} \sum_{\sigma=0}^{d_{\chi}} \mathcal{G}_{k, \gamma, \delta, \sigma, \alpha, \beta, j} \omega^\gamma r^{\delta} \chi^{\sigma} \partial_{r}^{\alpha} \partial_{\chi}^{\beta} h_j = 0 \,,
\end{split}
\end{equation}
where $\sum_{\alpha, \beta = 0}^{\alpha + \beta \leq 3}$ is a summation starting from $\alpha=0$ and $\beta=0$ up to $\alpha+\beta=3$ for all non-negative $\alpha$ and $\beta$, while $\mathcal{G}_{k, \gamma, \delta, \sigma, \alpha, \beta, j}$ is a complex function of $M$, $m$, and $a$ only.
The constants $d_r$ and $d_\chi$ are the degree of $r$ and $\chi$ of the coefficient of a given term in the linearized Einstein equations respectively, which depend on the specific equation we are looking at and can thus be thought of to be dependent on the summation indices $(\alpha,\, \beta, \, k, \, j)$. 
When factorizing each of the linearized Einstein equations to obtain the common denominator, there can be prefactors, such as  powers of $1-\chi^2$, $\Delta$ and $\Sigma$, that contain no metric perturbation functions and are nonzero except at $r = r_+$, and $\chi = \pm 1$. 
Since these common factors are never zero in the computational domain (except at the boundaries), we will divide by them to simplify the equations and improve the numerical stability of the linearized Einstein equations.

Equation~\eqref{eq:pertFE-1} represents a system of coupled, two-dimensional, third-order partial differential equations. Note that the perturbed field equations for the even perturbations are at most second order, whereas for odd perturbations the system of equations is at most third order, because of the $\partial_\chi h_k$ terms (with $h_k \in \{h_5,h_6\}$) in the metric decomposition of Eq.~\eqref{eq:odd}.
Each of the partial differential equations of Eq.~\eqref{eq:pertFE-1} consists of at least several thousand of terms. 
Moreover, the largest modulus of the numerical coefficients in $\mathcal{G}_{k, \gamma, \delta, \sigma, \alpha, \beta, j}$ of different equations can differ by $\sim$ 10 orders of magnitude, and the modulus of the coefficient of different terms of the same equation can also vary across $\sim 20$ orders of magnitude. 
To prevent overflow, we normalize every partial differential equation such that the largest modulus of the coefficient of each equation is one, which again, is allowed because of the homogeneous nature of the linearized Einstein equations.

Let us now derive the partial differential equations that $u_k(r, \chi)$ satisfy by substituting Eq.~\eqref{eq:radspec} into Eq.~\eqref{eq:pertFE-1}. 
Since the radial derivatives of the asymptotic factor can always be expressed as 
\begin{equation}
\frac{d^{\alpha}}{d r^{\alpha}} A(r) = \text{rational function} \times A(r), 
\end{equation}
after substituting Eq.~\eqref{eq:radspec} into Eq.~\eqref{eq:pertFE-1}, we can factorize out the $A(r)$, along with other common factors like $\Delta$ or the common denominator, for each equation. 
Since $A(r)$ and the other common factors are nonzero within the computational domain, we similarly remove them, leaving us with the following partial differential equations of $u_k(r, \chi)$, 
\begin{equation}\label{eq:pertFE-1u}
\begin{split}
& \sum_{j=1}^{6} \sum_{\alpha, \beta = 0}^{\alpha + \beta \leq 3} \sum_{\gamma=0}^{2} \sum_{\delta=0}^{\tilde{d}_{r}} \sum_{\sigma=0}^{\tilde{d}_{\chi}} \tilde{\mathcal{G}}_{k, \gamma, \delta, \sigma, \alpha, \beta, j} \omega^\gamma r^{\delta} \chi^{\sigma} \partial_{r}^{\alpha} \partial_{\chi}^{\beta} u_j = 0. 
\end{split}
\end{equation}
Here, $\tilde{\mathcal{G}}_{k, \gamma, \delta, \sigma, \alpha, \beta, j}$, is another complex function that depends on $M, m$ and $a$ but not on $r$ or $\chi$. 
$\tilde{d}_r$ and $\tilde{d}_\chi$ are, respectively, the degree of $r$ and $\chi$ of the coefficient of a given term in the equations of $u_k(r, \chi)$, which depend $(\alpha,\, \beta, \, k, \, j)$. 

We can now substitute the truncated spectral expansion of the metric perturbation functions (Eq.~\eqref{eq:spectral_decoposition_factorized}) into Eq.~\eqref{eq:pertFE-1u} to transform the latter into a system of linear algebraic equations. 
Since $r$ is a rational function of $z$ (see Eq.~\eqref{eq:z}), and by the chain rule of differentiation, 
\begin{equation}
\frac{\partial}{\partial r} = - \frac{(1+z)^2}{2 r_+} \frac{\partial}{\partial z}, 
\end{equation}
the coefficient functions of the linearized equations of $u_k$ must also be a rational function of $z$. 
Therefore, when we substitute Eq.~\eqref{eq:radspec} into Eq.~\eqref{eq:pertFE-1}, 
we can factorize the $k$-th partial differential equation as
\begin{equation}\label{eq:system_3}
\begin{split}
& \sum_{j=1}^{6} \sum_{\alpha, \beta = 0}^{\alpha + \beta \leq 3} \sum_{\gamma=0}^{2} \sum_{\delta=0}^{d_{z}} \sum_{\sigma=0}^{d_{\chi}} \mathcal{K}_{k, \gamma, \delta, \sigma, \alpha, \beta, j} \omega^\gamma z^{\delta} \chi^{\sigma} \partial_{z}^{\alpha} \partial_{\chi}^{\beta} u_j = 0 \,, 
\end{split}
\end{equation}
where $d_z$ and $d_{\chi}$ are the degree of $z$ and $\chi$ of the coefficient of the partial derivative $\partial_{z}^{\alpha} \partial_{\chi}^{\beta} \{...\} $ in the equations respectively, while $\mathcal{K}_{k, \alpha, \beta, \gamma, \delta, \sigma, j}$ are $k$ complex functions of $M, m$, $a$, $\rho_H^{(k)}$ and $\rho_{\infty}^{(k)}$ only (for every value of the summing indices $ \alpha, \beta, \gamma, \delta, \sigma$, and $j$). 

We conclude this subsection by pointing out that the linearized field equations in modified gravity theories can similarly be cast in the form of Eqs.~\eqref{eq:pertFE-1} or~\eqref{eq:system_3}, with the coefficient function of the partial derivatives of the metric perturbation variables as a polynomial functions of $r$ and $\chi$. 
In theories such as dynamical Chern-Simons gravity or Einstein-dilaton-Gauss-Bonnet gravity, BHs often couple to a scalar field. 
Schematically, the field equations linearized around these beyond-GR BHs take the form 
\begin{equation}\label{eq:beyond_GR_linearized_field_equation}
\begin{split}
\left[R_{\mu}{}^{\nu}\right]^{(1)} & = A_{\mu}{}^{\nu} (h_{\alpha \beta}, \Phi), \\
\Box \Phi (x^{\mu}) & = A_{\Phi} (h_{\alpha \beta}, \Phi), 
\end{split}
\end{equation}
where $\Box$ is the d'Alembertian operator defined with respect to the background BH spacetime, $\Phi$ is the scalar field to which the BH couples, and $ A_{\mu}{}^{\nu} (h_{\alpha \beta}, \Phi)$ and $A_{\Phi} (h_{\alpha \beta}, \Phi)$ are additional terms that depend linearly on $h_{\mu \nu}$ and $\Phi$. 
The METRICS approach can straightforwardly accommodate the scalar field by labeling $h_7 = \Phi $, changing the upper limit of $j$ to 7, and adding the scalar field equation to the linearized tensor equations. 
The field equations in these modified gravity theories involve only the derivatives of the Riemann or Ricci curvature tensors , or the Chirstoffel connections, or their products of integer power (see, e.g., \cite{Jackiw:2003pm, Alexander:2009tp, Yunes:2009hc} for the field equations of dynamical Chern-Simons gravity and for \cite{QNM_EdGB_01, QNM_EdGB_02, QNM_EdGB_03} Einstein-dilaton-Gauss-Bonnet gravity). 
This amounts to increasing the upper limit of $\gamma$ in Eqs.~\eqref{eq:pertFE-1} and~\eqref{eq:system_3}. 
In different modified gravity theories, BH solutions and the coupling scalar field are usually constructed in terms of rational functions, either numerically (e.g. \cite{Sullivan:2019vyi, Sullivan:2020zpf, McNees:2015srl}) or analytically via a perturbative scheme (see e.g. \cite{Yunes:2009hc, Maselli:2015tta, Cano_Ruiperez_2019}).
For the BHs whose space-time is not constructed in terms of rational functions, such as black holes with matter \cite{Nampalliwar:2021tyz, Daghigh:2022pcr}, we can still accurately approximate the space-time using the spectral functions by their completeness property and exponential convergence, upon suotable compactification(s). 
Thus, for a wide range of BH models, Eq.~\eqref{eq:beyond_GR_linearized_field_equation} also consists of only rational functions, which can also be cast in the form of Eqs.~\eqref{eq:pertFE-1} or~\eqref{eq:system_3} upon factorization. 
The additional coefficient functions (i.e. $\mathcal{G}$ and $\mathcal{K}$) can also be straightforwardly read from $A_{\mu}{}^{\nu} (h_{\alpha \beta}, \Phi)$ and the equation $\Box \Phi (x^{\mu}) = A_{\Phi} (h_{\alpha \beta}, \Phi)$.
The generality of Eq.~\eqref{eq:pertFE-1} and~\eqref{eq:system_3} therefore makes the METRICS approach extremely adaptable to compute the QNM frequencies of BHs outside GR or with matter, as long as the ingoing and outgoing principal null directions can be defined and found. 

\subsection{Converting the linearized Einstein equations into algebraic equations}
\label{sec:Conversion}

Let us now convert the linearized Einstein equations into an algebraic system of equations through use of our spectral expansion. We first substitute the truncated spectral expansion of the $u_k$ functions into Eq.~\eqref{eq:system_3}, 
\begin{equation}\label{eq:system_4}
\begin{split}
& \sum_{j=1}^{6} \sum_{\alpha, \beta = 0}^{\alpha + \beta \leq 3} \sum_{\gamma=0}^{2} \sum_{\delta=0}^{d_{z}} \sum_{\sigma=0}^{d_{\chi}} \mathcal{K}_{k, \gamma, \delta, \sigma, \alpha, \beta, j} \omega^\gamma z^{\delta} \chi^{\sigma} \\
& \quad \quad \quad \times \partial_{z}^{\alpha} \partial_{\chi}^{\beta} \Bigg\{\sum_{n=0}^{\mathcal{N}_z} \sum_{\ell=|m|}^{\mathcal{N}_{\chi}+|m|} v_j^{n \ell} T_{n}(z) P^{|m|}_{\ell}(\chi) \Bigg\} = 0.  
\end{split}
\end{equation}
These equations can be further simplified by using the defining equations for the Chebyshev polynomials and associated Legendre polynomials, namely
\begin{equation}
\begin{split}
\frac{d^2 T_n}{d z^2} & = \frac{1}{1-z^2} \left( z \frac{d T_n}{dz} - n^2 T_n \right), \\
\frac{d^2 P_{\ell}^{|m|}}{d \chi^2} & = \frac{1}{1-\chi^2} \Big( 2 \chi \frac{d P_{\ell}^{|m|}}{d \chi} - \ell(\ell+1) P_{\ell}^{|m|} \\
& \quad \quad \quad \quad \quad - \frac{m^2}{1-\chi^2} P_{\ell}^{|m|} \Big). 
\end{split}
\end{equation}
These equations allow us to pull out more factors of $1-\chi^2, 1-z$ or $1+z$, further simplifying Eq.~\eqref{eq:system_3}. 

We then express the left-hand side of Eq.~\eqref{eq:system_3} as a linear combination of the Chebyshev and associated Legendre polynomials, 
\begin{equation}\label{eq:elliptic_eqn_v2}
\sum_{n=0}^{\mathcal{N}_z} \sum_{\ell=|m|}^{\mathcal{N}_{\chi}+|m|} w_k^{n \ell} T_{n}(z) P^{|m|}_{\ell}(\chi) = 0\,, 
\end{equation}
where $w_k^{n \ell}$ are independent of $z$ and $\chi$, but depend on $M$, $a$, $n$, $\ell$, $m$, and\footnote{Recall that $n$ and $\ell$ here do not denote the overtone and azimuthal mode number of the QNM frequency. Rather, $n$ and $\ell$ are the order of the Chebyshev and the degree of the associated Legendre polynomials.} $\omega$, and $k \in (1,10)$. 
To satisfy the linearized Einstein equations, which are homogenous, we must then have that $ w_k^{n \ell} = 0 $ for every $k, n$ and $\ell$ by the orthogonality of $T_{n}(z)$ and of $P^{|m|}_{\ell}(\chi)$.
Inspecting Eqs.~\eqref{eq:system_3} and ~\eqref{eq:system_4}, we notice that $w_k^{n \ell}$ depends on $v_k^{n \ell}$ linearly, because
\begin{equation}
\label{eq:pertFE-2}
\begin{split}
w_k^{n \ell} = \sum_{j=1}^{6} \sum_{n'=0}^{\mathcal{N}_z} \sum_{\ell'=|m|}^{\mathcal{N}_{\chi}+|m|} \left[ \mathbb{D}_{n \ell, n' \ell'}(\omega) \right]_{kj} v_j^{n' \ell'} = 0, 
\end{split}
\end{equation}
where $\mathbb{D}_{n \ell, n' \ell'}(\omega)$ are $10 \times 6$ matrices, whose elements can be at most quadratic polynomials in $\omega$ and can be obtained by evaluating the inner product given by Eq. (41) of \cite{Chung:2023zdq}.  

Let us now introduce some new notations to simplify the expression of the equations we will have to solve. If we introduce the following vector notations \cite{Chung:2023zdq}, 
\begin{equation}\label{eq:vector_arrangement}
\begin{split}
& \textbf{v}_{n \ell} = \left( v_1^{n \ell}, v_2^{n \ell}, v_3^{n \ell}, v_4^{n \ell}, v_5^{n \ell}, v_6^{n \ell} \right)^{\rm T} \,, \\
& \textbf{w}_{n \ell} = \left( w_1^{n \ell}, w_2^{n \ell}, ..., w_{10}^{n \ell} \right)^{\rm T} \,,
\end{split}
\end{equation}
then Eq.~\eqref{eq:pertFE-2} can be written as, 
\begin{equation}\label{eq:vector_equations}
\textbf{w}_{n \ell} = \sum_{n'=0}^{\mathcal{N}_z} \sum_{\ell'=|m|}^{\mathcal{N}_{\chi}+|m|} \mathbb{D}_{n \ell, n' \ell'}(\omega) \textbf{v}_{n' \ell'} = 0\,,
\end{equation}
where the $\mathbb{D}_{n \ell, n' \ell'}$ matrix is now ``dotted'' into our new vector $\textbf{v}_{n' \ell'}$. 
Let us further define a vector $\textbf{v}$ and $\textbf{w}$, which, respectively, store all $\textbf{v}_{n \ell}$ and $\textbf{w}_{n \ell}$, as
\begin{equation}
\begin{split}
\textbf{v} & = \left\{ \textbf{v}_{00}^{\rm T}, \textbf{v}_{01}^{\rm T}, ..., \textbf{v}_{0 \mathcal{N}_{\chi}}^{\rm T}, ..., \textbf{v}_{1 \mathcal{N}_{\chi}}^{\rm T}, ...,  \textbf{v}_{(\mathcal{N}_{z}+1)(\mathcal{N}_{\chi}+1)}^{\rm T} \right\}^{\rm T}, \\
\textbf{w} & = \left\{ \textbf{w}_{00}^{\rm T}, \textbf{w}_{01}^{\rm T}, ..., \textbf{w}_{0 \mathcal{N}_{\chi}}^{\rm T}, ..., \textbf{w}_{1 \mathcal{N}_{\chi}}^{\rm T}, ...,  \textbf{w}_{(\mathcal{N}_{z}+1)(\mathcal{N}_{\chi}+1)}^{\rm T} \right\}^{\rm T}. 
\end{split}
\end{equation}
Note that $\textbf{v}$ is a $6(\mathcal{N}_{z}+1)(\mathcal{N}_{\chi}+1)$-vector, whereas $\textbf{w}$ is a $10(\mathcal{N}_{z}+1)(\mathcal{N}_{\chi}+1)$-vector.
Then, Eq.~\eqref{eq:pertFE-2} can be more compactly written as 
\begin{equation}\label{eq:augmented_matrix_01}
\begin{split}
\textbf{w} = \tilde{\mathbb{D}} (\omega) \textbf{v} = \left[ \tilde{\mathbb{D}}_0 + \tilde{\mathbb{D}}_1 \omega + \tilde{\mathbb{D}}_2 \omega^2 \right] \textbf{v} = \textbf{0} \,,
\end{split}
\end{equation}
where the $\tilde{\mathbb{D}}_{0,1,2}$ matrices are constant, $ 10 (\mathcal{N}_z + 1)(\mathcal{N}_{\chi} + 1) \times 6 (\mathcal{N}_z + 1)(\mathcal{N}_{\chi} + 1) $ rectangular matrices.
The QNM frequencies of a Kerr BH correspond to the $\omega$ such that Eq.~\eqref{eq:augmented_matrix_01} admits a nontrivial solution $\textbf{v}$. 

\section{Numerical methodology to solve the eigenvalue problem in the METRICS Approach}
\label{sec:extraction}

Unlike in the Schwarzschild BH case, to accurately compute the QNM frequencies of a Kerr BH we have to simultaneously solve all ten linearized Einstein equations (Eq.~\eqref{eq:system_3}) for the six finite correction functions ($u_k(z, \chi)$). 
In practice, this can be achieved through a Newton-Raphson method whose details, advantages and numerical implementations will be explained next.

\subsection{Newton-Raphson method applied to QNM frequencies in the METRICS approach}
\label{sec:newton_method}

Equation~\eqref{eq:augmented_matrix_01} is a linear, homogenous equation for $\textbf{v}$, which creates two difficulties in computing the QNM frequencies. 
First, Eq.~\eqref{eq:augmented_matrix_01} admits a trivial solution, to which the Newton-Raphson algorithm may accidentally converge.
Second, Eq.~\eqref{eq:augmented_matrix_01} is linear in $\textbf{v}$ so the solution is not unique, i.e.,~if $\textbf{v}$ is a solution, then $c \textbf{v}$, where $c$ is a constant, is also a solution of Eq.~\eqref{eq:augmented_matrix_01}. 

To remedy these two difficulties, we separately consider perturbations ``led'' by the polar and axial sectors, i.e., perturbations that have a larger amplitude in a particular parity. 
To obtain perturbations led by a given parity, we set the initial guess of the Newton-Raphson method for the metric perturbations of the opposite parity to zero, and a spectral coefficient of the leading parity to unity.
Without loss of generality, for polar-led perturbations, we start with an initial guess of $h_{1 \leq k \leq 4} = 0$ and $v_{k=5}^{n=0,\ell=|m|} = 1$; for axial-led perturbations, we start with $h_{k =5, 6} = 0$ and $v_{k=1}^{n=0,\ell=|m|} = 1$. 
Upon iteration of the Newton-Raphson method, however, the mode that was initialized to zero will be driven away from zero because our initial guess does not satisfy the linearized Einstein equations.

The isospectrality of the Teukolsky equation, and thus of the Einstein equations, however, guarantees that both the polar- and axial-led modes will have the same QNM frequencies. In other words, since the spectrum is isopectral, any QNM $q$ must have the same $\omega_{q}^{\rm (P)}$ and $\omega_{q}^{\rm (A)}$. Therefore, we expect that if we consider polar-led or axial-led perturbations, we ought to find the same QNM frequencies. 
As a side note, if considering a modified gravity theory where isospectrality is broken, we need the initial guesses to be such that purely axial or purely polar perturbations are excited, which we will explore in future work.

Let us then focus our discussion on polar-led perturbations only, keeping in mind that the procedures applied below apply to the axial-led perturbations similarly. 
As mentioned, for polar-led perturbations, our initial guess of $\textbf{v}$ is
\begin{equation}\label{eq:polar_convention}
v_{k=1}^{n=0,\ell=|m|} = 1\,,
\end{equation}
which breaks the linear-scaling invariance of Eq.~\eqref{eq:augmented_matrix_01}, so that, for a given $\omega$, Eq.~\eqref{eq:augmented_matrix_01} admits a nontrivial solution. 
With this convention in place, we now have $ 10 (\mathcal{N}_{z}+1)(\mathcal{N}_{\chi}+1) $ equations for $6 (\mathcal{N}_{z}+1)(\mathcal{N}_{\chi}+1)$ unknowns, which are the remaining mode components $v^{n \neq 0, \ell \neq 0}_{k\neq 1}$ and $\omega$. 
We denote all these unknowns by a $6 (\mathcal{N}_{z}+1)(\mathcal{N}_{\chi}+1)$-vector $\textbf{x}$, 
\begin{equation}
\textbf{x}^{\rm (P)} = \left\{ v^{n \neq 0, \ell \neq 0}_{k\neq 1}, \omega^{\rm (P)} \right\}^{\rm T}. 
\end{equation}
Symbolically, we denote Eq.~\eqref{eq:augmented_matrix_01} when $v_{k=1}^{n=0,\ell=|m|} = 1$ by 
\begin{equation} \label{eq:vector_equations_w_polar_convention}
f^{\rm (P)}_k = \left[ \sum_{j=1}^{6} \sum_{n'=0}^{\mathcal{N}_z} \sum_{\ell'=|m|}^{\mathcal{N}_{\chi}+|m|} \left[ \mathbb{D}_{n \ell, n' \ell'}(\omega) \right]_{kj} v_j^{n' \ell'} \right]_{v_{k=1}^{n=0,\ell=|m|} = 1} \hspace{-1cm}=0, 
\end{equation}
where the subscript ``(P)'' reminds us that we are focusing on the polar-led perturbations, and the bracket notation $[X]_{v_{k=1}^{n=0,\ell=|m|} = 1}$ implies that $X$ is evaluated with the condition $v_{k=1}^{n=0,\ell=|m|} = 1$ of Eq.~\eqref{eq:polar_convention} enforced. The quantity $f^{\rm (P)}$ can be thought of as a vector-valued function of the vector $\textbf{x}^{\rm (P)}$, 
\begin{equation}
\textbf{f}^{\rm (P)} (\textbf{x}) = [ f^{\rm (P)}_{k}(\textbf{x}) ]_{k=1, 2, ..., 10 (\mathcal{N}_{z}+1)(\mathcal{N}_{\chi}+1)}. 
\end{equation}

There are several approaches to solve Eq.~\eqref{eq:vector_equations_w_polar_convention}, such as gradient descent, variable projection \cite{varpro}, or Newton-Raphson (see e.g.~\cite{Dias:2015nua}). 
In this work, we solve Eq.~\eqref{eq:vector_equations_w_polar_convention} through the Newton-Raphson method. 
To devise the Newton-Raphson iteration procedure, we consider the infinitesimal differences of $\textbf{f}^{\rm (P)} $ due to an infinitesimal displacement of $\textbf{x}^{\rm (P)}$ from an initial solution, 
\begin{equation}
d \textbf{f} = \textbf{J} \cdot d \textbf{x},  
\end{equation}
where $\textbf{J}$ is the $10 (\mathcal{N}_{z}+1)(\mathcal{N}_{\chi}+1) \times 6 (\mathcal{N}_{z}+1)(\mathcal{N}_{\chi}+1)$ Jacobian matrix, whose $(i,j)$th element is given by 
\begin{equation}
[\textbf{J}]_{ij} = \frac{\partial f_i}{\partial [\textbf{x}]_{j}} \Bigg|_{\textbf{x} = \textbf{x}_{(n)}}, 
\end{equation}
and we have dropped the superscript $\rm (P)$ for clarity. 
Then we write 
\begin{equation}
\begin{split}
d \textbf{x} & = \textbf{x}_{n+1} - \textbf{x}_{n}, \\
d \textbf{f} & = \textbf{f}(\textbf{x}_{n+1}) - \textbf{f}(\textbf{x}_{n}) \approx - \textbf{f} (\textbf{x}_{n}),   \\
\end{split}
\end{equation}
where we have approximated $\textbf{f}(\textbf{x}_{n+1})$ by 0 because $\textbf{x}_{n+1}$ should be a better guess than $\textbf{x}_{n}$. 
Hence, given a guess $\textbf{x}_{n}$, we can update it by solving 
\begin{equation}
\textbf{J} \cdot \left( \textbf{x}_{n+1} - \textbf{x}_{n} \right) = - \textbf{f}(\textbf{x}_{n}), 
\end{equation}
which can readily be done using the Moore-Penrose inverse of $\textbf{J}$
\begin{equation}
\textbf{x}_{n+1} = \textbf{x}_{n} - \textbf{J}^{-1} \cdot \textbf{f}(\textbf{x}_{n}). 
\end{equation}
Here $\textbf{J}^{-1}$ denotes the Moore-Penrose inverse of $\textbf{J} $, which is a matrix of $6 (\mathcal{N}_{z}+1)(\mathcal{N}_{\chi}+1) \times 10 (\mathcal{N}_{z}+1)(\mathcal{N}_{\chi}+1)$. 
As the residual vector, $\textbf{f}(\textbf{x}_{n})$ is a $10 (\mathcal{N}_{z}+1)(\mathcal{N}_{\chi}+1)$-vector, $\textbf{J}^{-1} \cdot \textbf{f}(\textbf{x}_{n})$ gives a $6 (\mathcal{N}_{z}+1)(\mathcal{N}_{\chi}+1)$-vector, which is of the same length as $\textbf{x}_{n}$.

Formally, the Newton-Raphson method does not provide an exact solution, but rather it yields an approximate numerical solution that satisfies the equations to a specified error tolerance. The iterative method then ends when the error tolerance is reached. In this work, we terminate the iterations when
\begin{equation}\label{eq:tolerance_error}
\| \textbf{f}(\textbf{x}_{n}) \|_{2} < \epsilon, 
\end{equation}
where $ \| \textbf{f}(\textbf{x}_{n}) \|_{2} $ is the $2$ norm of the residual vector $\textbf{f}(\textbf{x}_{n})$ and $\epsilon$ is the error tolerance. 
Henceforth, we will set $\epsilon = 10^{-7}$.
If $\| \textbf{f}(\textbf{x}_{n}) \|_{2} = 0$, all ten linearized Einstein equations are satisfied exactly. 

For the axial-led perturbations, we set
\begin{equation}\label{eq:axial_convention}
v_{k=5}^{n=0,\ell=|m|} = 1,
\end{equation}
and we then solve the following system of algebraic equations
\begin{equation} \label{eq:vector_equations_w_axial_convention}
f^{\rm (A)}_k = \left[ \sum_{j=1}^{6} \sum_{n'=0}^{\mathcal{N}_z} \sum_{\ell'=|m|}^{\mathcal{N}_{\chi}+|m|} \left[ \mathbb{D}_{n \ell, n' \ell'}(\omega) \right]_{kj} v_j^{n' \ell'} \right]_{v_{k=5}^{n=0,\ell=|m|} = 1} \hspace{-1cm}= 0.
\end{equation}
We solve this equation using the Newton-Raphson algorithm explained above with $\textbf{f}^{\rm (A)} (\textbf{x}) = \{ f_{k}^{\rm (A)} (\textbf{x}) \}$ and $\textbf{x} = \textbf{x}^{\rm (A)} = \left\{ v^{n \neq 0, \ell \neq 0}_{k\neq 5}, \omega^{\rm (A)} \right\}^{\rm T}$.
When solving for axial-led modes, we will employ the same tolerance $\epsilon$ as when solving for polar-led modes.

\subsection{Advantages of implementing the Newton-Raphson method}

Now that we have explained the details of the Newton-Raphson algorithm, we can discuss its advantages over the previous implementation of \cite{Chung:2023zdq}.
First, the version of the Newton-Raphson method described above allows us to solve all ten linearized Einstein equations simultaneously.
Previously, in \cite{Chung:2023zdq}, we solved only six (out of the ten) linearized Einstein equations to compute the QNM frequencies. 
In principle, the resulting $\omega$ and $\textbf{v}$ obtained by solving only six equations may not satisfy the remaining four (whether they do or not depend on which six equations are chosen), but formally, physically viable metric perturbations should satisfy all linearized Einstein equations. 
The Newton-Raphson method described above ensures that the $\omega$ and $\textbf{v}$ correspond to metric perturbations that satisfy all ten linearized Einstein equations, as well as the four linearized Bianchi identities, the latter of which implies the conversation of energy and momentum of the metric perturbations. 
Thus, the Newton-Raphson method guarantees that the resulting metric perturbations are physically viable. 

Second, the Newton-Raphson method avoids spurious eigenvalues, i.e., solutions to the quadratic eigenvalue problem that do not actually correspond to true QNM frequencies. 
This method focuses on one QNM at a time, and it only explores the physically motivated regions in the complex plane. As a result, the Newton-Raphson method saves computational time and effort. 
In contrast, the method in \cite{Chung:2023zdq} transforms the quadratic eigenvalue problem into a linearized generalized eigenvalue problem, which leads to many spurious eigenvalues for both numerical and physical reasons. Numerically, the finite precision to which one computes the matrices $\mathbb{D}_{i=0,1,2}$ and the numerical instability introduced by the transformation contributes to the existence of spurious eigenvalues. 
Physically, under-resolving the physics in the solution (for example, because of truncation of the approximate spectral expansion), and misrepresenting the physics in the solution (for example, due to the use of a leading-order asymptotic expansion in the resummation of the controlling factor) can also lead to spurious eigenvalues.
Thus, with the implementation in \cite{Chung:2023zdq}, one needs to identify and separate the physical eigenvalues from the spurious ones. 
On the other hand, the Newton-Raphson algorithm focuses on one QNM at a time, avoiding the identification problem entirely. 

Third, enforcing the conventions in Eq.~\eqref{eq:polar_convention} ensures that the eigenvector at different spectral orders corresponds to the same metric perturbations, instead of being arbitrarily scaled by a multiplicative constant. 
Solving the whole set of algebraic equations in this way allows us to reconstruct metric perturbations, while simultaneously computing the QNM frequency. 
This is drastically more convenient than reconstructing the metric perturbations using the CKK formalism applied to the Teukolsky solution. 
By keeping track of the changes in $\textbf{x}$,  we can also monitor the numerical stability and terminate the computation when a certain accuracy is reached. 
All these improvements make studying the gravitational perturbations of a rotating BH with the METRICS approach more efficient.

Finally, the Newton-Raphson method keeps the dimension of the coefficient matrix unchanged if we apply the spectral method to other modified gravity theories. For example, in dynamical Chern-Simons gravity, the linearized field equations usually involve third-order time derivatives of the metric \cite{PratikRingdown,Wagle:2021tam}. 
In the frequency domain and upon spectral expansion, the linearized field equations involving these higher-order time derivatives result in an eigenvalue problem of higher-than-second degree in the QNM frequency. 
If one transforms this eigenvalue problem into a linear generalized eigenvalue problem, as done in~\cite{Chung:2023zdq}, then the order (size) of the resulting matrices will be greatly increased, which requires more memory to store them and more computational resource to compute the eigenvalues. 
However, using the Newton-Raphson algorithm, no augmentation of the matrix is needed, and the dimension of the matrix that we need to handle remains unchanged, making the spectral method more easily generalizable to modified gravity.

\subsection{Numerical implementation}
\label{sec:setup}

In this subsection, we explain the setup for the numerical computations of the QNM frequencies with the Newton-Raphson method.
To simplify our discussion, we assume $\mathcal{N}_z = \mathcal{N}_{\chi} = N$. We denote the axial-led frequency of a given mode $q$ computed using $N \times N$ spectral functions by $\omega_q^{\text{(A)}} (N)$ and its polar-led counterpart by $\omega_q^{\text{(P)}} (N)$. 
Similarly, we denote the metric perturbations reconstructed using $N \times N$ spectral functions by, 
\begin{equation}
u_k(z, \chi, N) = \sum_{n=0}^{N} \sum_{\ell=|m|}^{N+|m|} v_k^{n \ell}(N) T_{n}(z) P^{|m|}_{\ell}(\chi),
\end{equation}
where $v_k^{n \ell}(N)$ are the mode coefficients obtained using $N \times N$ spectral functions. With this at hand, $\tilde{\mathbb{D}}(\omega)$ becomes a $10(N+1)^2 \times 6 (N+1)^2 $ rectangular matrix and $\textbf{x}$ is a $6 (N+1)^2$ vector.

Let us begin by discussing the initial guess of the QNM frequency to initialize the Newton-Raphson method. 
We will here choose the first two significant figures of the real and imaginary parts of the known QNM frequencies of Kerr BHs at the corresponding value of the dimensionless spin $a$ as the initial QNM frequency guess for most of our computations. 
Doing so results in a speed up of our calculations, which will allow us to explore the spectrum more efficiently. 
This choice, however, does not affect sensitively the METRICS frequencies we will obtain, as we will show in more detail in Sec.~\ref{sec:sensitivity_to_initial_guess}. 
For example, we obtain very similar METRICS frequencies when we use just the first significant digit of the known QNM frequencies of Kerr BHs as our initial guess.
Moreover, if we did not know the QNM frequencies of the Kerr BH at all, we would still be able to implement an initial guess as follows. 
Since the Kerr metric is a continuous function of $a$, the QNM frequency should also depend on $a$ continuously. 
Thus, we could start by considering a BH with small spin $a_1$ and use the QNM frequencies of the Schwarzschild BH as the initial guess. 
Once the Newton-Raphson method converges to a QNM frequency for a BH with spin $a_1$, we could use this value as the initial guess for a BH with a slightly larger spin. 
Repeating this process iteratively, we would be able to build a tower of QNM frequencies with guesses from the previous iteration. 

This argument can be generalized to BHs in modified gravity. For theories that can be treated as deformations from GR, continuously parameterized by coupling constants and with a smooth GR limit, we expect the QNM spectra to branch out from the Kerr spectra continuously in the complex plane. Therefore, we should be able to obtain the modified gravity QNM frequencies for small deformations from GR using the Kerr frequencies as our initial guess. For larger deformations, we can then take the QNM frequencies of the small coupling constants as the initial guess, and continue recursively.

The numerical implementation of the inverse of the Jacobian matrix requires detailed discussion. 
We compute the Moore-Penrose inverse using the built-in \texttt{PseudoInverse} function of \texttt{Mathematica} with double precision. 
The Moore-Penrose inverse and all quantities in the subsequent computations are stored and computed with double precision. 
We checked that increasing the precision limit of the calculations does not significantly affect our results.  

As mentioned before, the termination criterion of the Newton-Raphson method is controlled by a tolerance, which we set here to $\epsilon = 10^{-7}$. 
That is, we iterate the Newton-Raphson method until the $2$-norm of the residual vector is below this tolerance. 
We find empirically that the $2$-norm of the residual vector of converged solutions is usually $10^{-9} - 10^{-8}$ by the end of the iteration process (see Fig.~\ref{fig:residual_N}), and further reducing the tolerance has no significant effects on the final results. 
Moreover, when the Newton-Raphson algorithm converges, it usually converges within $\sim 10$ steps. 

As implied in the previous paragraph, there are indeed cases for which the Newton-Raphson method does not converge with the numerical settings described above. 
This is particularly the case when $N$ is too small to accurately represent the solution to the linearized Einstein field equations.
While we expect that the linearized Einstein equations should always admit a nontrivial solution corresponding to a QNM, mathematically, algebraic equations resulting from the spectral expansion of the linearized Einstein equations may only admit the trivial solution. 
At best, the resulting algebraic equations admit an approximated nontrivial solution, whose residual ($\textbf{f}(\textbf{x})$) is small. 
If the minimal residual at a given QNM frequency is larger than the prescribed tolerance, the iterator becomes trapped in an infinite cycle and the iterations never converge. 
To curtail these pathological behavior of our Newton-Raphson implementation, we perform at most 20 iterations for a given $N$, and store the resulting $\omega$ at the 20th iteration. 
This choice of maximal iteration number does not affect our computations, because only the iterations of small $N$ are affected; in general, we will select $\omega$s computed with a large enough $N$ such that the iterations converge within 10 steps.

\section{QNM Frequencies from the METRICS approach}
\label{sec:results}

In this section, we apply the procedures developed in the last section to compute the dominant mode ($n=0, l=2, m=2$, the so-called ``022" mode) frequency of the Kerr BHs. 
Since the $022$ mode is the QNM that usually dominates the ringdown phase of compact binary coalescence \cite{LIGOScientific:2021sio, pyRing_02, pyRing_03}, we focus on it in this paper, although the METRICS approach can be applied to any $nlm$ mode.
We will first show the numerical results concerning the 022-mode frequency of the Kerr BH of $a=0.1$ and study its properties. 
Then, we will apply the METRICS approach to compute the fundamental frequency of more rapidly rotating Kerr BHs with $a\leq 0.95$. 

\subsection{022-mode frequency of Kerr BHs with dimensionless spin $a=0.1$}
\label{sec:a_0.1}

\begin{figure*}[htp!]
\centering  
\subfloat{\includegraphics[width=0.47\linewidth]{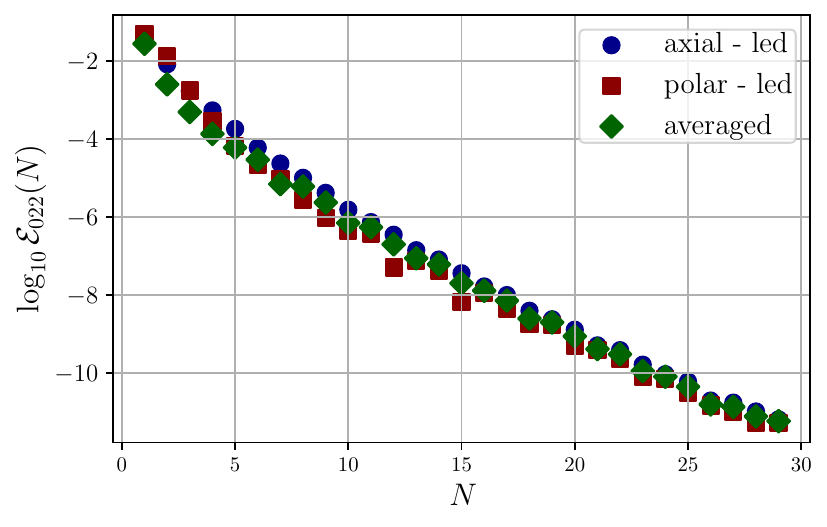}}
\subfloat{\includegraphics[width=0.47\linewidth]{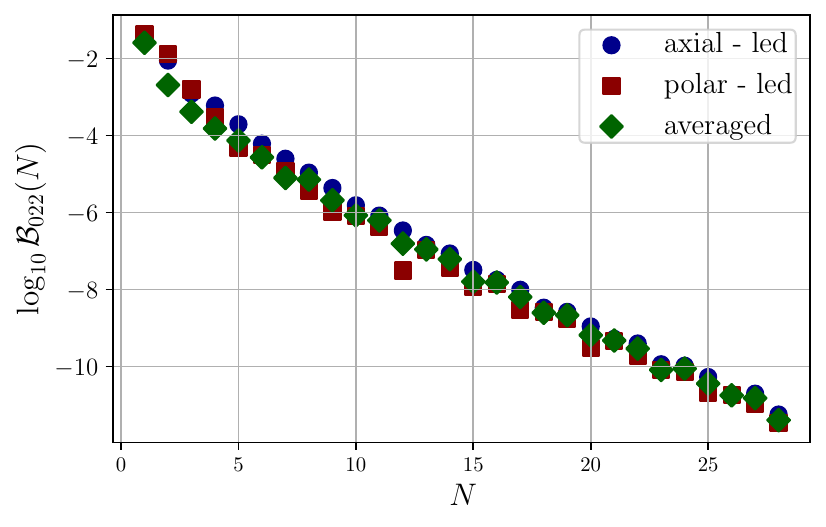}}
\caption{Absolute error (left) and backward modulus difference (right) of the axial-led (blue circles), polar-led (red squares) and averaged 022-mode frequency (green diamonds) of a Kerr BH with dimensionless spin $a=0.1$, computed using the METRICS approach as a function of the spectral order $N$. 
At all spectral orders, the initial guess for the Newton-Raphson method is the first two significant digits of the known, 022-mode frequency. 
Observe that the base-10 logarithm of the error and backward modulus difference decreases approximately linearly with spectral order, as expected from the exponential convergence of spectral expansions. 
}
\label{fig:a_0.1}
\end{figure*}

Let us begin by computing $\omega_{022}$ of a Kerr BH of dimensionless spin $a=0.1$. 
Although this BH has a small rotation rate, its numerical results reflect many common features that we also observe when we study more rapidly rotating Kerr BHs. 
Before proceeding to quantify the properties of the numerical results, we define an error measure, which is the absolute error between the 022-mode frequency computed via the METRICS approach (using $N$ Chebyshev and associated Legendre polynomials) and that computed by solving the Teukolsky equation using Leaver's continued fraction method, 
\begin{equation}\label{eq:Error_rho_1}
\begin{split}
& \mathcal{E}^{\rm(A/P)}_{022}(N) = |\omega^{\rm (A/P)}_{022}(N) - \omega_{022}(\rm L)|, 
\end{split}
\end{equation}
where the superscript (A) and (P) denote whether the mode studied is axial- or polar-led, and $\omega_{022}(\rm L)$ is the 022-mode frequency computed using Leaver's method. 

The left panel of Fig.~\ref{fig:a_0.1} shows the base-10 logarithms of $\mathcal{E}^{\rm(A/P)}_{022}(N) $ of the Kerr BH of $a=0.1$ as a function of $N$. 
From Fig.~\ref{fig:a_0.1}, we observe two features. 
First, $\log_{10} \mathcal{E}^{\rm(A/P)}_{022}(N)$ decreases with $N$ approximately linearly for $N\in[1, 29]$. 
This implies that both the axial- and polar-led, 022, METRICS QNM frequencies converge to the Leaver frequency exponentially with $N$, which is consistent with the exponential-convergence property of spectral expansions \cite{boyd2013chebyshev}. 
The exponential convergence can also be quantified by computing the backward modulus difference of $\omega_{022}^{\rm (P/A)}$, 
\begin{equation}\label{eq:backward_modulus_diff}
\mathcal{B}_{022}^{\rm (P/A)}(N) = |\omega_{022}^{\rm (P/A)}(N+1) - \omega_{022}^{\rm (P/A)}(N)|. 
\end{equation}
The right panel of Fig.~\ref{fig:a_0.1} shows the base-10 logarithms of $\mathcal{B}^{\rm(A/P)}_{022}(N) $ as a function of $N$. 
Observe that $\log_{10} \mathcal{B}^{\rm(A/P)}_{022}(N)$ varies with $N$ approximately linearly, consistent with the pattern of $\log_{10} \mathcal{E}^{\rm(A/P)}_{022}(N) $ and the expected exponential convergence of the spectral expansion. 
The monotonically decreasing property of $\mathcal{B}_{022}(N)$ suggests that we should always select the $\omega_{022}(N) $ of the largest computed $N$ for the range of $N$s we investigated.

\begin{figure}[htp!]
\centering  
\subfloat{\includegraphics[width=\columnwidth]{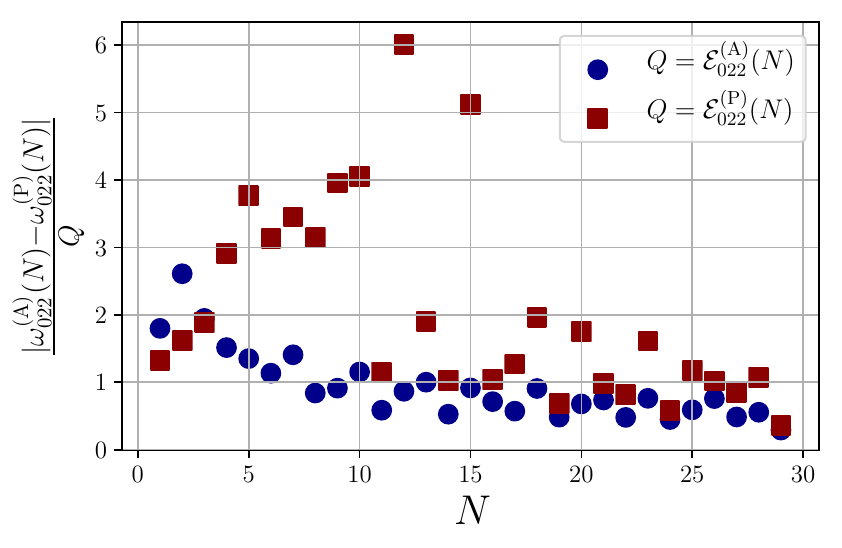}}
\caption{Ratio of $|\omega^{\rm(A)}_{022}(N) - \omega^{\rm(P)}_{022}(N)|$ to $\mathcal{E}^{\rm(A)}_{022}(N)$ (blue circles) and to $\mathcal{E}^{\rm(P)}_{022}(N)$ (red squares) as a function of $N$. 
At $N = 29$, $|\omega^{\rm(A)}_{022}(N) - \omega^{\rm(P)}_{022}(N)|$ is smaller than both $\mathcal{E}^{\rm(P)}_{022}(N)$ and $\mathcal{E}^{\rm(P)}_{022}(N)$, indicating that, although $\omega^{\rm(A)}_{022}(N)$ and $\omega^{\rm(P)}_{022}(N)$ are not formally identical due to numerical errors, they are still consistent with each other, and thus, with the expected isospectrality of the QNM frequencies of BHs in GR. 
}
\label{fig:a_0.1_isospectrality}
\end{figure}

The second observation we can draw from the left panel of Fig.~\ref{fig:a_0.1} is that, at a given $N$, $\mathcal{E}^{\rm(A)}_{022}(N) $ and $\mathcal{E}^{\rm(P)}_{022}(N) $ are very close to each other. 
Since both  $\mathcal{E}^{\rm(A)}_{022}(N) $ and $\mathcal{E}^{\rm(P)}_{022}(N) $ decrease exponentially, so does the difference $|\omega^{\rm(A)}_{022}(N)-\omega^{\rm(P)}_{022}(N)|$.
This implies that $\omega^{\rm(A)}_{022}(N) $ and $\omega^{\rm(P)}_{022}(N) $ are also very close to each other, which is seemingly consistent with the expected isospectrality of QNM frequencies of gravitational perturbations of BHs in GR. 
A savvy observer, however, will notice that $\omega^{\rm(A)}_{022}(N) $ and $\omega^{\rm(P)}_{022}(N) $ are not identical, and their difference,  $|\omega^{\rm(A)}_{022}(N) - \omega^{\rm(P)}_{022}(N)|$, is therefore a measure of the error of the Newton-Raphson algorithm.
Nonetheless, if $|\omega^{\rm(A)}_{022}(N) - \omega^{\rm(P)}_{022}(N)|$ is smaller than the absolute errors $\mathcal{E}^{\rm(A)}_{022}(N)$ and $\mathcal{E}^{\rm(P)}_{022}(N)$, the METRICS 022-mode frequency is still consistent with isospectrality. 

Figure~\ref{fig:a_0.1_isospectrality} shows the ratio of $|\omega^{\rm(A)}_{022}(N) - \omega^{\rm(P)}_{022}(N)|$ to $\mathcal{E}^{\rm(A)}_{022}(N)$ (blue circles) and to $\mathcal{E}^{\rm(P)}_{022}(N)$ (red squares) as a function of $N$. 
Observe that, for small $N$, $|\omega^{\rm(A)}_{022}(N) - \omega^{\rm(P)}_{022}(N)|$ can be larger than $\mathcal{E}^{\rm(A)}_{022}(N)$ or $\mathcal{E}^{\rm(P)}_{022}(N)$, which is reasonable because when $N$ is small, the METRICS frequency is not very accurate, and the solutions we find will not necessarily respect isospectrality. 
Importantly, however, as $N$ increases and becomes large, and in particular, when $N=29$, $|\omega^{\rm(A)}_{022}(N) - \omega^{\rm(P)}_{022}(N)|$ is smaller than either $\mathcal{E}^{\rm(A)}_{022}(N)$ or $\mathcal{E}^{\rm(P)}_{022}(N)$. This indicates that the differences between the axial- and polar-led frequencies are within the absolute error of the frequency, thus confirming the expected isospectrality of GR. 
We have also checked that $|\omega^{\rm(A)}_{022}(N) - \omega^{\rm(P)}_{022}(N)|$ is smaller than the absolute error for larger values of $a$. 
Since $\omega^{\rm(A)}_{022}(N) $ and $\omega^{\rm(P)}_{022}(N) $ are almost the same, from here on, we take their average and define this to be the METRICS 022-mode frequency, 
\begin{equation}
\omega_{022} (N) = \frac{1}{2} \left( \omega^{\rm(A)}_{022}(N) + \omega^{\rm(P)}_{022}(N)\right). 
\end{equation}
Quantities associated with the averaged frequency, such as the absolute error and backward modulus difference, are labeled as ``averaged" in the figures throughout the paper. 

\begin{figure}[tp!]
\centering  
\subfloat{\includegraphics[width=\columnwidth]{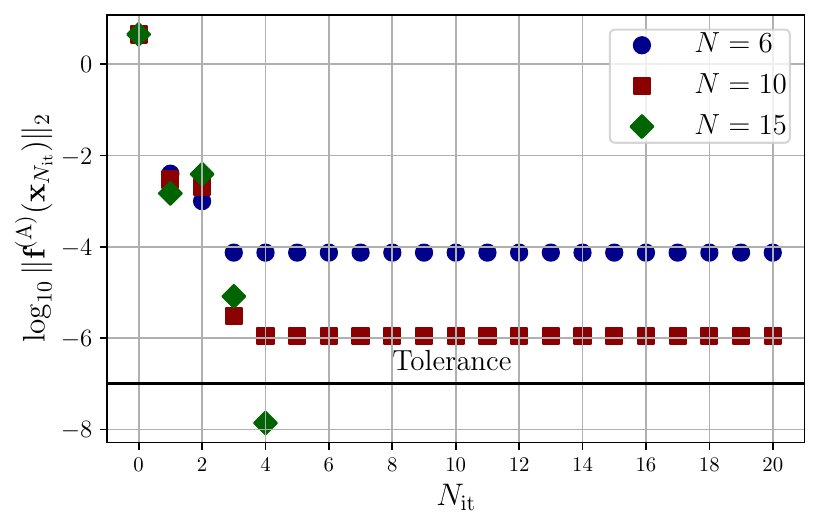}}
\caption{Residual of the algebraic equations at various, fixed spectral orders $N$, but as a function of the iteration number $N_{\rm it}$ of the Newton-Raphson method used to compute the QNM frequency of an $a=0.1$ Kerr BH, where the horizontal line marks the error tolerance.
Only the residual of the algebraic equations related to the axial-led frequency is shown, because that related to the polar-led one is quantitatively similar.
Observe that, for all spectral orders, the residual decreases as $N_{\rm it}$ increases. 
For small spectral orders (i.e. $N = 6 $ and 10), the residual reaches a plateau as $N_{\rm it} \sim 3~\rm or~4$, which is smaller for the $N=10$ case than for the $N=6$ case.
This is because the algebraic equations cannot faithfully approximate the linearized Einstein equations if the spectral order is too small. 
When the spectral order is sufficiently large (e.g. $N=15$), the residual can become smaller than the tolerance error, at which point the iterations are terminated. 
}
 \label{fig:residual_N}
\end{figure}

To ensure the frequency and eigenvector that we solve for using the METRICS approach satisfy all ten linearized Einstein equations, we monitor the residual $\| \textbf{f}(\textbf{x}_{N_{it}}) \|_2$ as the iterations progress at every spectral order $N$.  
As pointed out in Sec.~\ref{sec:setup}, the algebraic equations [Eqs.~\eqref{eq:vector_equations_w_polar_convention} and~\eqref{eq:vector_equations_w_axial_convention}] are just a spectral approximation of the linearized Einstein equations, and thus, they may not admit an exact solution. 
Nonetheless, if the algebraic equations are a good approximation to the linearized Einstein equations, we still expect to be able to obtain a solution $\textbf{x}$ such that $\| \textbf{f}(\textbf{x})\|_2 $ is extremely small as compared to the initial guess.
Figure~\ref{fig:residual_N} shows $\log_{10} \| \textbf{f}^{\rm (A)}(\textbf{x}_{N_{\rm it}})\|_2$ as a function of the iteration number ($N_{\rm it}$) at different spectral orders, with tolerance error ($\epsilon = 10^{-7}$) marked by a solid, horizontal, black line. 
We show only the residual of the axial-led computations because the residual of the polar-led computations is quantitatively similar. 
Observe that for, all $N$, the initial residual is the same because the same initial guess (cf. Eq.~\eqref{eq:axial_convention}) is used, but as the iterations progress, the residual decreases. 
For small spectral orders (i.e. $N = 6 $ and $10$), the residual reaches a plateau as $N_{\rm it} $ reaches 3 or 4, with the minimum residual deceasing for the larger $N$ case.
The plateau and the decrease of the minimal residual is because the algebraic equations cannot faithfully approximate the solution to the linearized Einstein equations if the spectral order is too small. 
If we approximate the solution to the linearized Einstein equations with a sufficiently high spectral order, the minimum residual decreases monotonically with iteration number, until it becomes smaller than the tolerance. 
The iterations are then terminated, as in the $N=15$ (green diamonds) case. 
Observe that, at $N=15$, the final residual is about ten orders of magnitude smaller than the residual of the initial guess, indicating that the algebraic equations, and thus all the ten linearized Einstein equations, are well satisfied numerically. 

\subsection{022-mode frequency of Kerr BHs with moderate dimensionless spins}

\begin{figure*}[htp!]
\centering  
\subfloat{\includegraphics[width=0.47\linewidth]{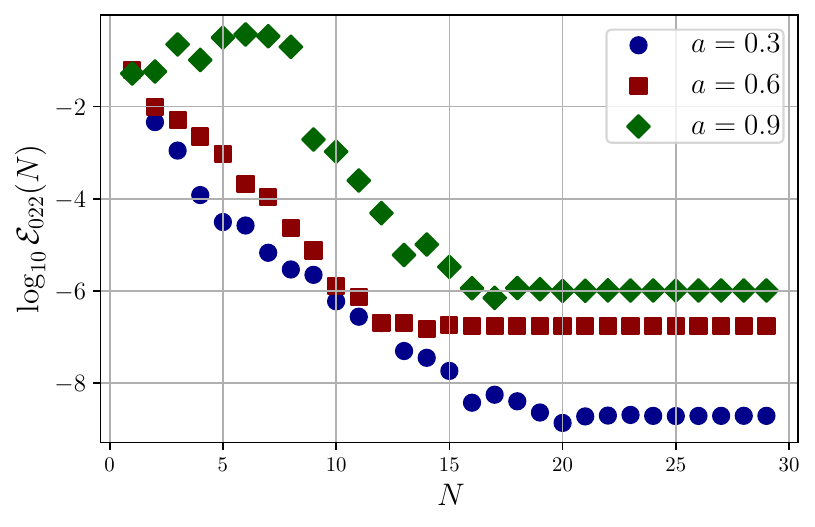}}
\subfloat{\includegraphics[width=0.47\linewidth]{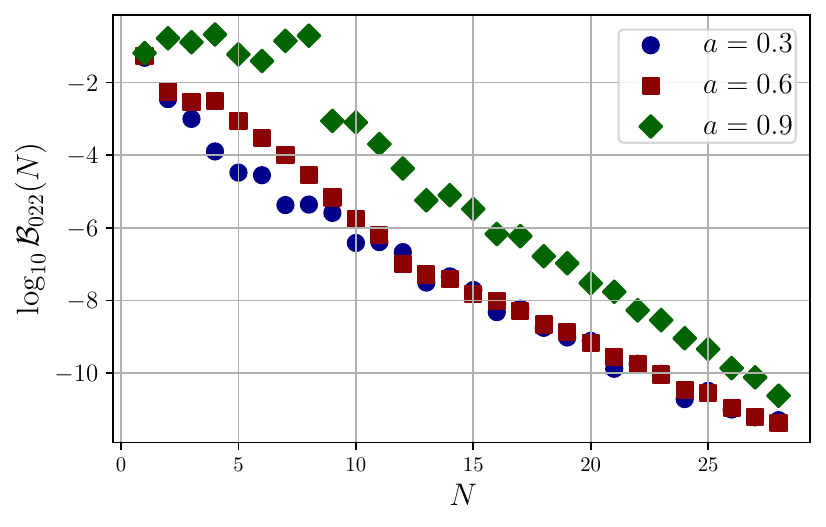}}
\caption{The absolute error (left) and backward modulus difference (right) of the averaged 022-mode frequency of a Kerr BH of $a=0.3$ (blue circles), 0.6 (red squares), and 0.9 (green diamond) computed using the spectral method as a function of the spectral order $N$.
At all spectral orders, the first two significant digits of the known 022-mode frequency are used as the initial guess to initiate the Newton-Raphson iterations. 
We observe that $\log_{10} \mathcal{E}_{022} (N) $ of different dimensionless spins first decreases approximately linearly with $N$. 
When $N$ reaches an $a$-dependent value, $\log_{10} \mathcal{E}_{022} (N) $ for different dimensionless spins reaches an $a$-dependent plateaus. 
Since $\log_{10} \mathcal{B}_{022}(N)$ decreases approximately linearly with $N$ for $N \in [1, 29]$, the spectral 022-mode frequency is exponentially converging to a frequency which shows a tiny deviation of modulus of $\sim 10^{-6}$ from the Leaver frequency. 
This is actually reasonable because we are solving the linearized Einstein equations transformed from the original ones using the asymptotic factor and the compactified coordinate.
Through the transformation, some physics is inevitably misrepresented, leading to the deviation of the METRICS frequency from the Leaver frequency. 
}
\label{fig:multi_as}
\end{figure*}

We now apply the METRICS approach to compute the fundamental mode frequency of more rapidly rotating BHs. 
The left panel of Fig.~\ref{fig:multi_as} shows the $\mathcal{E}_{022}(N) $ of the averaged 022-mode frequency of a Kerr BH with $a = 0.3$ (blue circles), 0.6 (red squares) and 0.9 (green diamonds).
We show only the averaged frequency because the axial- and polar-led frequencies are very close to each other when $N$ is large, as pointed out in Sec.~\ref{sec:a_0.1}. 
Observe that, for all dimensionless spins shown, $\log_{10} \mathcal{E}_{022}(N)$ first decreases approximately linearly with $N$. 
This feature was also found in the $a=0.1$ case, and it is consistent with the exponential convergence of spectral expansions. 
Nonetheless, as $N$ reaches a certain value that depends on $a$, the absolute error reaches an $a$-dependent plateau (which is unrelated to the plateau of the residual we studied as a function of iteration number in the Newton-Raphson method).  

To understand the nature of these plateaus, we compute $\mathcal{B}_{022}(N)$ for different dimensionless spins, which is shown in the right panel of Fig.~\ref{fig:multi_as}. 
As $\log_{10} \mathcal{B}_{022}(N)$ decreases approximately linearly with $N$ throughout $N \in [1, 29]$, we conclude that $\omega_{022}$ at different dimensionless spins converges exponentially to a value that is different from the correct one, as given by Leaver's, continuous fraction method.
In other words, the METRICS QNM frequencies show a tiny deviation from the correct value, whose modulus is $\sim 10^{-6}$ from the correct frequency. 
One possible reason of such a deviation may be the breakdown of the validity of separating $u_k(r, \chi)$ into a product of a function of $r$ and a function of $\chi$ at high dimensionless spin. 
The metric components of the Kerr BH spacetime (cf. Eq.~\eqref{eq:metric}) are not separable in such a product decomposition, since they contain powers of $r^2 + a^2 \chi^2$ in denominators, which becomes increasingly manifest as $a$ increases.
In view of this nonseparability, perhaps a more sophisticated function of $\chi$ is needed as a basis of the spectral expansion.
Another possible reason for the deviation may lie in the inaccuracy of the asymptotic factor $A(r)$. 
In the limit $r\rightarrow r_+$, $A(r)$ describes the diverging behavior of the metric perturbations at the event horizon, which depends on $a$. 
However, $A(r)$ captures only the leading-order asymptotic behavior of the metric perturbations. 
To improve the accuracy of the method, perhaps subleading orders in the asymptotic behavior of $A(r)$ also need to be included. 
All of the above calls for further investigations if one wishes to improve the implementation of the METRICS approach. 
With that in mind, nonetheless, the modulus of the deviation is of the order of $10^{-6}$, which is small compared to the existing and foreseeable QNM frequency measurement using ground-based detectors (see, e.g., \cite{LIGOScientific:2021sio}). 
Therefore, we leave such improvements to the METRICS approach to future studies. 

\subsection{022-mode frequency of Kerr BHs with dimensionless spins $a \leq 0.95$}

Using the METRICS approach, we now present results for the fundamental mode frequency of Kerr BHs with $a \leq 0.95$. 
Table~\ref{tab:results} lists the real (second column) and imaginary parts (third column) of the averaged 022-mode frequency at the corresponding $a$ (left-most column). 
All METRICS QNM frequencies are computed using 30 Chebyshev and associated Legendre polynomials. 
We show only the first ten digits of the METRICS QNM frequencies because the backward modulus difference is $\sim 10^{-10}$ at $N=29$. 
For reference, the fourth and fifth columns of Table~\ref{tab:results} respectively list the real and imaginary part of fundamental frequency computed with Leaver's method at the corresponding spin. 
More specifically, $\omega(\rm L)$ is the fundamental frequency obtained by solving the Teukolsky equation using Leaver's method of continued fractions with 150 terms in the fraction; keeping this large number of terms, we find that $\omega(\rm L)$ is not changed up to the $16\textrm{th}$ decimal at machine precision.
To make sure that the QNM frequencies (and the corresponding eigenvector) satisfy all ten linearized Einstein equations, the right most two columns show the residual ratio $\mathcal{R}$, i.e.~the ratio between the residual when the iterations are terminated and that at the initial guess, namely 
\begin{equation}\label{eq:residual_ratio}
\mathcal{R} = \frac{\| \textbf{f}(\textbf{x}_{\rm terminated}) \|_{2} }{\| \textbf{f}(\textbf{x}_{1}) \|_{2} }. 
\end{equation}
If the resulting frequency and eigenvector satisfy all ten linearized Einstein equations well, then $\mathcal{R}$ should be close to zero, which is indeed the case for all dimensionless spins we calculated. 

\begin{table*}[htb]
\begin{tabular}{c|cc|cc|c|c}
\hline
$a$ & $\omega_{\rm Re} (\rm spectral)$ & $\omega_{\rm Im} (\rm spectral)$ & $\omega_{\rm Re} (\rm Leaver)$ & $\omega_{\rm Im} (\rm Leaver)$ & axial-led $\mathcal{R}$ & axial-led $\mathcal{R}$ \\ \hline
0.005 & $0.3743023147$ & $-0.0889522054$ & $0.374302314745705$ & $-0.0889522053640457$ & $8.02\times 10^{-11}$ & $1.49\times 10^{-13}$ \\
0.1 & $0.3870175384$ & $-0.0887056990$ & $0.3870175383645592$ & $-0.0887056990268991$ & $9.36\times 10^{-12}$ & $3.85\times 10^{-11}$ \\
0.2 & $0.4021453242$ & $-0.0883111662$ & $0.4021453241072112$ & $-0.08831116615465$ & $6.68\times 10^{-13}$ & $1.35\times 10^{-12}$ \\
0.3 & $0.4195266818$ & $-0.0877292719$ & $0.4195266799093153$ & $-0.0877292712328145$ & $4.85\times 10^{-9}$ & $2.38\times 10^{-9}$ \\
0.4 & $0.4398419217$ & $-0.0868819620$ & $0.439841909727434$ & $-0.0868819580547294$ & $9.32\times 10^{-11}$ & $4.34\times 10^{-11}$ \\
0.5 & $0.4641230260$ & $-0.0856388350$ & $0.4641229739649294$ & $-0.0856388194008764$ & $6.21\times 10^{-10}$ & $1.62\times 10^{-10}$ \\
0.6 & $0.4940447818$ & $-0.0837652022$ & $0.4940446109217166$ & $-0.0837651572095065$ & $1.33\times 10^{-9}$ & $1.59\times 10^{-10}$ \\
0.7 & $0.5326002436$ & $-0.0807928732$ & $0.5325997998444519$ & $-0.0807927741196761$ & $5.00\times 10^{-11}$ & $1.31\times 10^{-11}$ \\
0.8 & $0.5860169749$ & $-0.0756295524$ & $0.5860160981862801$ & $-0.07562938913772186$ & $1.89\times 10^{-9}$ & $1.36\times 10^{-9}$ \\
0.9 & $0.6716142721$ & $-0.0648692359$ & $0.671613259501218$ & $-0.06486906741255006$ & $3.31\times 10^{-11}$ & $1.01\times 10^{-11}$ \\
0.95 & $0.7463199985$ & $-0.0531490080$ & $0.7463194371599231$ & $-0.05314891507283093$ & $4.72\times 10^{-11}$ & $8.34\times 10^{-11}$ \\ \hline
\end{tabular}
\caption{
\label{tab:results}
Averaged 022-mode frequency (real part in the second column and imaginary part in the third column) computed using 30 Chebyshev and associated Legendre polynomials at different dimensionless spins $a$ (first column). 
At all dimensionless spins, the first two significant digits of the known 022-mode frequency are used as the initial guess to initiate the Newton-Raphson iterations. 
Only the first ten digits of the METRICS QNM frequencies are shown because the backward modulus difference of the 022-mode frequency is $\sim 10^{-10} $ if 30 Chebyshev and associated Legendre polynomials are used. 
For reference, we show the real and imaginary parts of frequencies computed with Leaver's method on the third and fourth columns, keeping up to 150 terms in the continuous fraction. 
The two right most columns show the ratio between the final residual and that at the initial guess (see Eq.~\eqref{eq:residual_ratio} for more details). 
}
\end{table*}

\begin{figure*}[tp!]
\centering  
\subfloat{\includegraphics[width=0.7\linewidth]{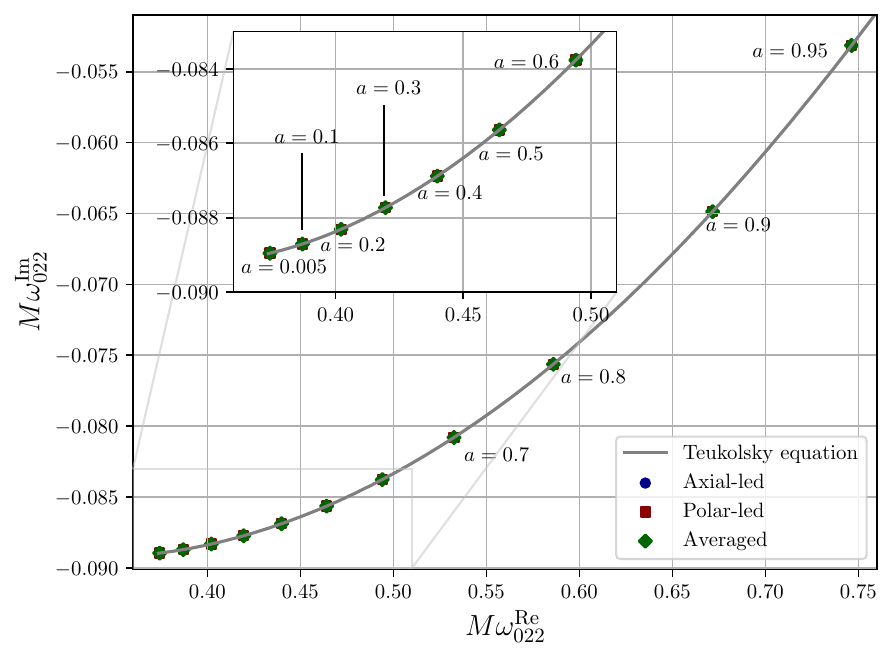}}
\caption{Axial-led, polar-led and averaged 022-mode frequency of a Kerr BH of different dimensionless spin, from $a=0.005$ to $a=0.95$, in the complex plane, computed with the METRICS approach. 
The solid line represents the frequencies computed by solving the Teukolsky equation using Leaver's method. 
The circles and squares show the values of the polar-led and axial-led QNM frequencies, while the stars show the averaged values. 
Observe that the METRICS 022-mode frequencies coincide almost exactly with those computed with Leaver's method, indicating that the spectral method is capable of accurately computing the fundamental frequencies of a generically rotating BH. 
}
 \label{fig:Kerr_022}
\end{figure*}

To visualize the above results, we present the axial-led (blue circles), polar-led (red squares), and averaged (black stars) 022-mode frequency computed using the spectral method for $0.005 \leq a \leq 0.95$ in Fig.~\ref{fig:Kerr_022}. 
For comparison, the fundamental frequencies computed with Leaver's method are shown as a function of $a$ with a solid gray line.
Figure~\ref{fig:Kerr_022} allows us to make several observations.
First, both the axial- and polar-led QNM frequencies are very close to the known values computed using Leaver's method at the corresponding spin, indicating that our spectral method can accurately compute the 022-mode frequency for rapidly rotating BHs.
Second, the axial- and polar-led QNM frequencies almost coincide with each other, once again indicating consistency with the isospectrality of gravitational QNM spectra of BHs in GR. 

Although the QNM frequencies computed by the spectral method in Fig.~\ref{fig:Kerr_022} appear to be consistent with the known values, we would still like to quantify the accuracy of our results. 
To this end, we will employ the absolute error measure of Eq.~\eqref{eq:Error_rho_1}, the residual ratio of Eq.~\eqref{eq:residual_ratio}, and a new measure: the relative fractional error in the real and imaginary parts of the polar-led, axial-led and the averaged spectral QNM frequencies relative to that computed with Leaver's method~\cite{Leaver:1985ax}, namely
    \begin{equation}\label{eq:relative_error}
    \begin{split}
    & \Delta^{\rm (P/A)}_{\rm Re / Im} = \left|1-\frac{\omega^{\rm (P/A)}_{\rm Re / Im} (\text{METRICS})}{\omega_{\rm Re / Im} (\text{L})}\right|\,, \\
    & \Delta_{\rm Re / Im} = \left|1-\frac{\omega_{\rm Re / Im} (\text{METRICS})}{\omega_{\rm Re / Im} (\text{L})}\right|\,, \\
    \end{split}
    \end{equation}
    where $\omega_{\rm Re / Im} (\text{METRICS})$ and $\omega_{\rm Re / Im} (\text{L})$ stand for the real and imaginary parts of $\omega(\text{METRICS})$ and $\omega (\text{L})$ respectively.

\begin{figure*}[htp!]
\centering  
\subfloat{\includegraphics[width=0.47\linewidth]{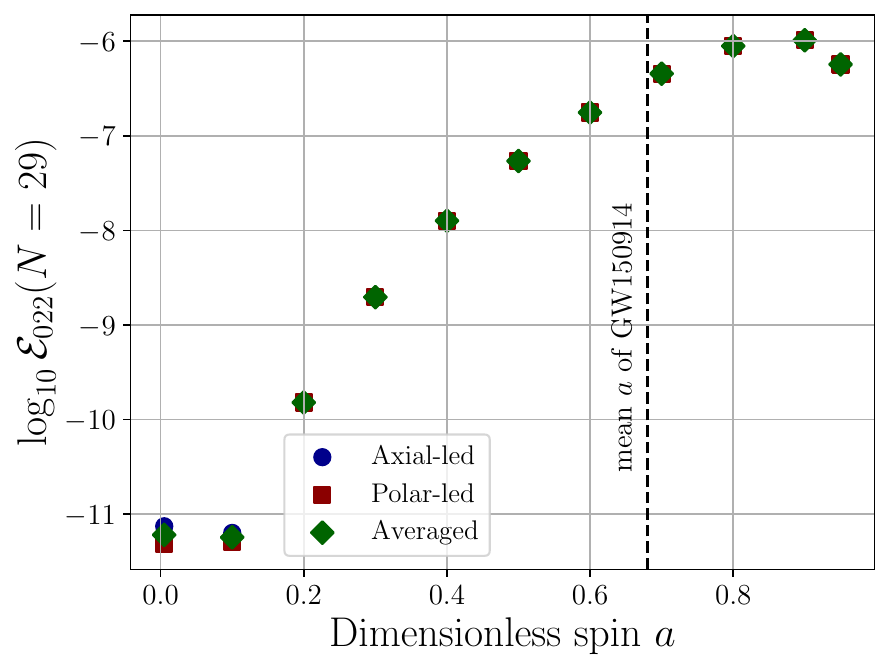}}
\subfloat{\includegraphics[width=0.47\linewidth]{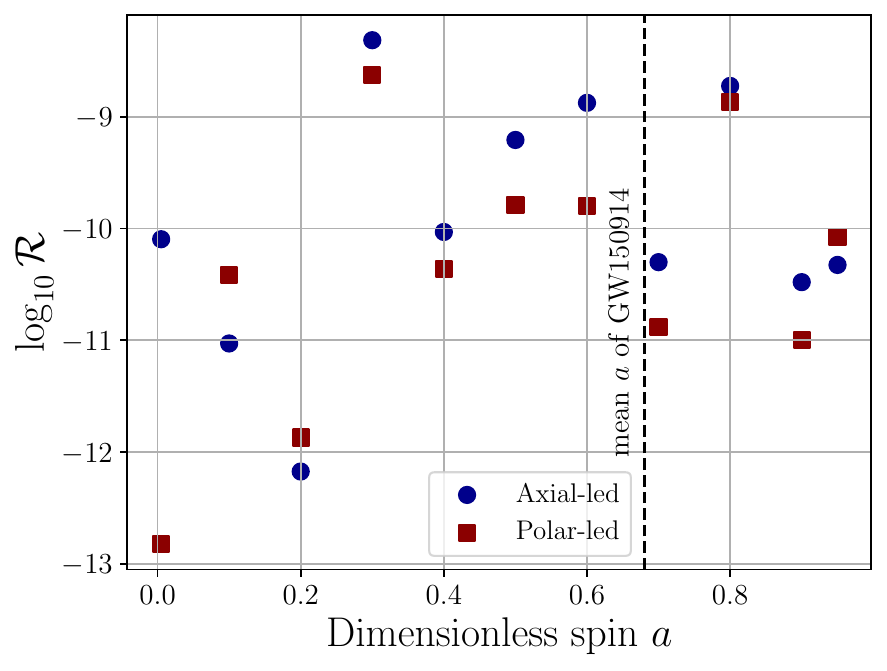}}
\qquad
\subfloat{\includegraphics[width=0.47\linewidth]{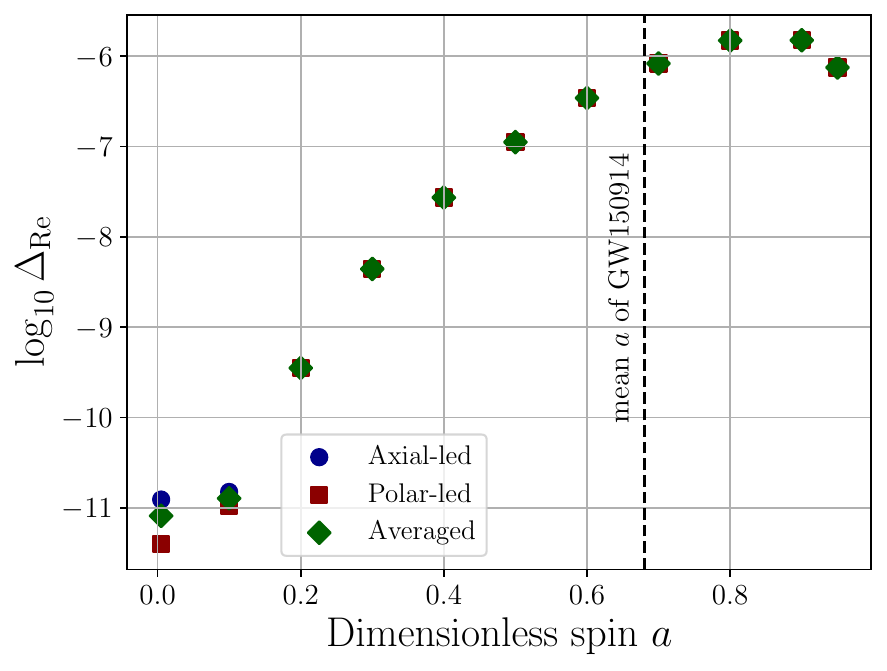}}
\subfloat{\includegraphics[width=0.47\linewidth]{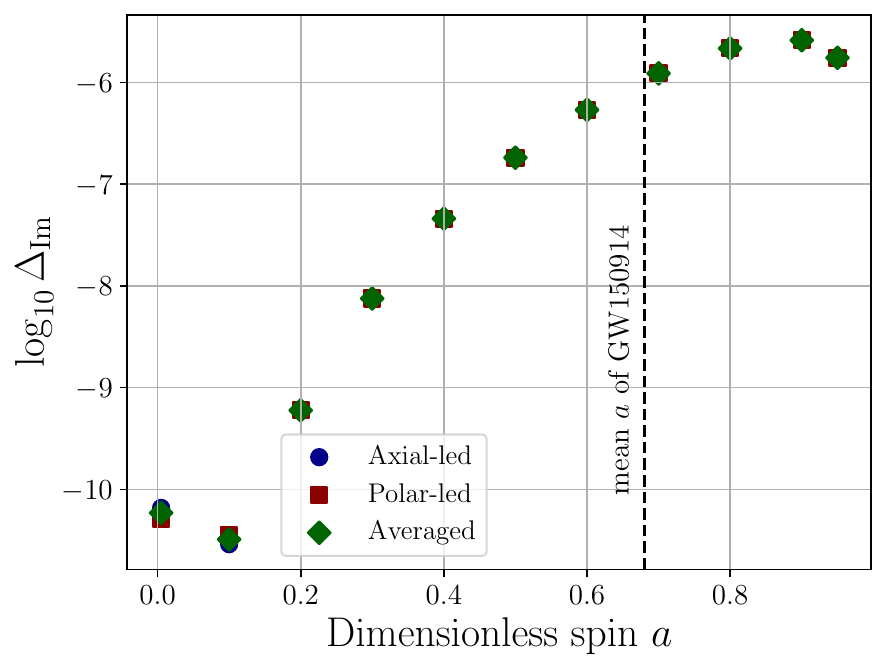}}
\caption{To gauge the accuracy of the spectral method, we compare the METRICS QNM frequencies ($\omega_\text{METRICS}$) to those computed through Leaver's method \cite{Leaver:1985ax} ($\omega_\text{L}$).
The top left panel shows the absolute error $\mathcal{E} = |\omega_\text{METRICS} - \omega_\text{L}|$.
The bottom left and right panels show the relative fractional error in the real ($\Delta^{\text{Re}} = \left|1-\omega^{\text{Re}}_\text{METRICS}/\omega^{\text{Re}}_\text{L}\right|$) and imaginary ($\Delta^{\text{Im}} = \left|1-\omega^{\text{Im}}_\text{METRICS}/\omega^{\text{Im}}_\text{L}\right|$) parts of the 022-mode frequency, respectively.
The top right panel shows the ratio between the residual of the linearized Einstein equation at the initial step and the final step of the Newton-Raphson iterations (see the main text for definition). 
In all panels, the dashed vertical line marks the mean value of the $a$ of the remnant BH of GW150914 reported in \cite{LIGOScientific:2016vlm}.
We observe that for the relative fractional error of the real and imaginary parts of the METRICS 022-mode frequency at all spin is significantly smaller than that obtained by combining all the LIGO-Virgo ringdown signals, which is $\sim 10^{-2}$ \cite{LIGOScientific:2021sio}. 
This suggests that the spectral 022-mode frequency is accurate enough to be applied for analyzing actual ringdown signals, where the 022 mode dominates.
}
\label{fig:Kerr_022_QNMF_error}
\end{figure*}

The top-left, bottom-left, and bottom-right panels of Fig.~\ref{fig:Kerr_022_QNMF_error} show the logarithmic (base 10) absolute error ($\mathcal{E}_{022}$), relative fractional errors in the real part ($\Delta_{\text{Re}}$), and the imaginary part ($\Delta_{\text{Im}}$) parts of the spectral 022-mode frequency for the axial-led (blue circles), polar-led (red squares), and averaged (black diamonds) frequencies, respectively, as a function of $a$.
Observe that, in general, all three errors increase steadily as $a$ increases.
Quantitatively, all errors first increase sharply from $\sim 10^{-10}$ to $10^{-6}$ as $a$ increases from 0 to $0.6$, and then they fluctuate around $10^{-6}$ from $a=0.6$ onward. This behavior reflects the deviation of the METRICS 022-mode frequency from the true frequency of rapidly rotating BHs, as shown by the error plateau in the left panel of Fig.~\ref{fig:multi_as}.

Crucially, the relative error of the METRICS frequency is about four orders of magnitude smaller than that obtained by combining all the LIGO-Virgo ringdown signals, which is $\sim 10^{-2}$ \cite{LIGOScientific:2021sio}, and is also significantly smaller than the (projected) relative fractional measurement uncertainty of combining $\sim 10^3$ ringdown signals detected with the next-generation detectors, which is  $\sim 10^{-4}$ \cite{Maselli:2023khq}.
This small relative fractional error of the METRICS frequency indicates that the METRICS frequencies are accurate enough to use in the analysis of existing and future ringdown signals detected by ground-based detectors. 

The top-right panel of Fig.~\ref{fig:Kerr_022_QNMF_error} shows the logarithmic (base 10) residual ratio ($\mathcal{R}$) as a function of $a$.
From this panel, we identify no significant correlation between $\mathcal{R}$ and $a$. 
Nonetheless, $\mathcal{R}$ for both leading parities is $\lesssim 10^{-9}$, which means that the resulting QNM frequency and eigenvector satisfy all ten linearized Einstein equations quite well. 

\section{Robustness of Quasinormal Frequency Extraction with the METRICS approach}
\label{sec:robustness}

In this section, we study the robustness of the calculations presented in the previous section. 
In particular, we first focus on exploring the geometry of the solution space by computing the residual as a function $\omega$, with the other components of $\textbf{x}$ fixed at the eigenvector obtained using the Newton-Raphson algorithm. 
We then check the robustness of the QNM frequency obtained using the Newton-Raphson algorithms by computing the fundamental frequency using different initial guesses. 
Finally, we study the effects of our choice of boundary conditions for the $\rho_{H,\infty}^{(k)}$ constants in the controlling factor of the spectral expansion.

\subsection{Search for spurious minima in solution space}

\begin{figure}[tp!]
\centering  
\subfloat{\includegraphics[width=\columnwidth]{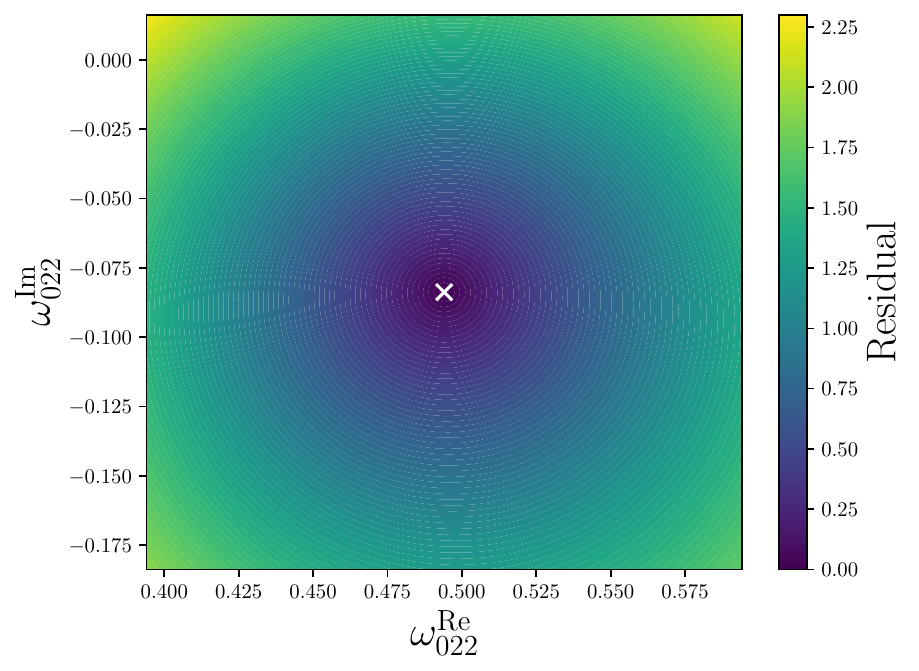}}
\caption{The residual, $\| \textbf{f} (\textbf{x})\|_{2}$ (see Eq.~\eqref{eq:tolerance_error} for definition), of computing the axial-led 022 frequency of a Kerr BH of $a=0.6$ and $N = 14$ as a function of $\omega$ in the complex plane.
The eigenvector $\textbf{v}$ is fixed at the one obtained using the Newton-Raphson iterations. 
The color plot covers a square of width of 0.2 centered at the Leaver 022-mode frequency, which is marked by the white cross. 
We observe that the residual is minimized at a position near the known 022-mode frequency, with no other local minima around.
This suggests the Newton-Raphson iterator is not likely to be attracted or trapped in a local minimum which does not correspond to a physical QNM frequency. 
The residual of the polar-led frequency is similar. 
}
 \label{fig:Heat_map}
\end{figure}

We first explore the robustness of the QNM frequency calculations by exploring if there exists other (unphysical) solutions to Eqs.~\eqref{eq:vector_equations_w_polar_convention} and~\eqref{eq:vector_equations_w_axial_convention} that are near the one we obtained by the Newton-Raphson iterations. 
We check this by computing the residual, $\| \textbf{f}(\textbf{x})\|_2$, as a function of $\omega$ with $\textbf{v}$ fixed at that found from the Newton-Raphson method. 
Figure~\ref{fig:Heat_map} shows a color plot of $\| \textbf{f}^{\rm (A)}(\textbf{x})\|_{2}$ as a function of $\omega$ when computing the axial-led frequency of a Kerr BH with dimensionless spin  $a=0.6$ and $N=14$ terms in the spectral expansion. 
The color plot covers a square of width 0.2, centered at the fundamental frequency obtained with Leaver's method for a Kerr BH with $a=0.6$ (marked by the white cross). 
We choose this spin because it corresponds to a rapidly rotating BH whose frequency does not require many spectral bases to be accurately computed (see. Fig.~\ref{fig:multi_as}). 
We show only the results of the axial-led perturbations because those concerning the polar-led perturbations are similar. 
Observe that the residual is minimized only at a position that is very close to the known Leaver frequency, with no local minima surrounding the correct fundamental frequency. This indicates that the Newton-Raphson method will not be attracted or trapped in local minima that do not correspond to a physical QNM frequency. 

\subsection{Sensitivity to initial guess}
\label{sec:sensitivity_to_initial_guess}

\begin{figure}[tp!]
\centering  
\subfloat{\includegraphics[width=\columnwidth]{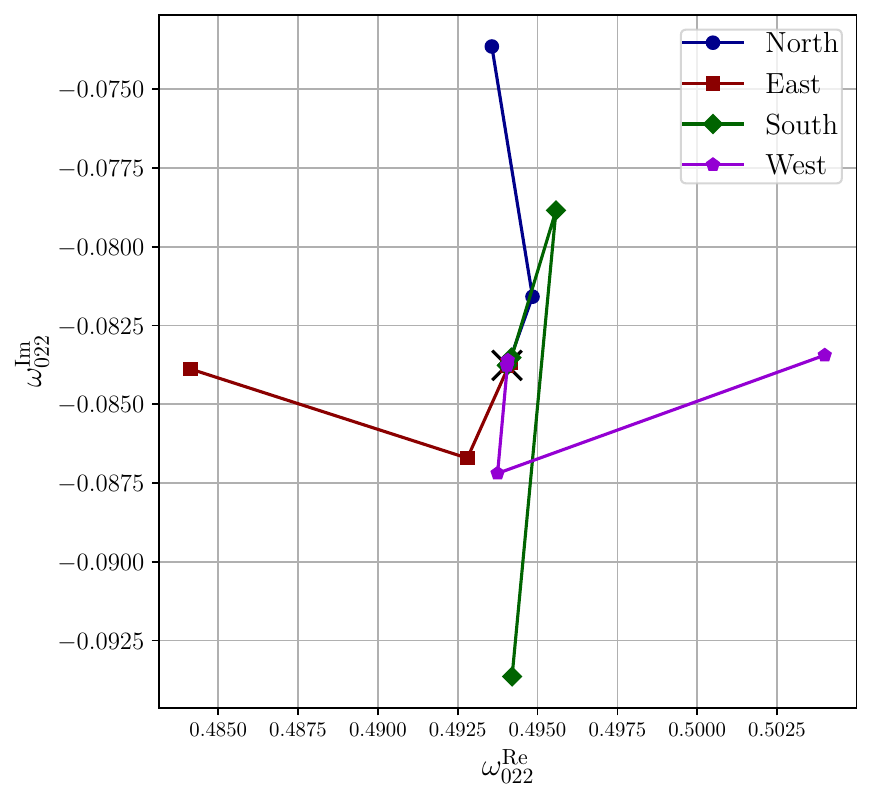}}
\caption{The trajectory in the complex plane traced by the Newton-Raphson iterator as we use differential initial guesses to compute the axial-led 022-mode frequency of a Kerr BH of $a=0.6$ of spectral order 14. 
We only show the trajectories of the axial-led frequencies because the trajectories of computing the polar-led frequency are similar. 
The different initial guesses are obtained by displacing the known 022-mode frequency (black cross) by 0.01 northward (blue circles), eastward (red squares), southward (green diamonds), and westward (violet pentagons). 
We observe that, within three iterations, the iterator has approached the known frequency very closely. 
We verify that starting from all initial guesses, the iterations reach the tolerance error of $10^{-7}$ in six iterations.
These results indicate that the QNM frequencies computed using our spectral method are robust against the choices of initial guesses. 
}
 \label{fig:NESW_test}
\end{figure}

Next, we check that the resulting QNM frequency does not depend on the initial guess sensitively. To study this, we displace the initial guess by 0.01 ``northward'', ``eastward'', ``southward'', and ``westward'' in the complex plane, relative to the fundamental frequency computed with Leaver's method. 
Figure~\ref{fig:NESW_test} shows the ``trajectory'' in the complex plane traced by the Newton-Raphson iterator for these different initial guesses, when computing the axial-led 022-mode frequency of a Kerr BH of $a=0.6$ with $N=14$ terms in the spectral expansion. 
The cross in the figure represents the Leaver 022-mode frequency for this Kerr BH. 
We only show the trajectories of the axial-led frequencies because the trajectories of the polar-led frequency are similar. 
Observe that, regardless of the direction of the displacement of the initial guess, the Newton-Raphson iterator approaches the known frequency within three iterations. 
Moreover, regardless of the initial guess chosen, the iterations reach the tolerance error of $10^{-7}$ within six iterations. 
The resulting absolute error of the axial-led frequencies is smaller than $10^{-6}$ in all cases. 
The results of this test indicate that the QNM frequencies computed using our spectral method are robust against the choice of initial guess, allowing us to accurately compute the QNM frequencies without having an accurate initial guess \textit{a priori}. 

\subsection{Effects of $\rho_H^{(k)}$ and $\rho_{\infty}^{(k)}$}

\begin{figure}[tp!]
\centering  
\subfloat{\includegraphics[width=\columnwidth]{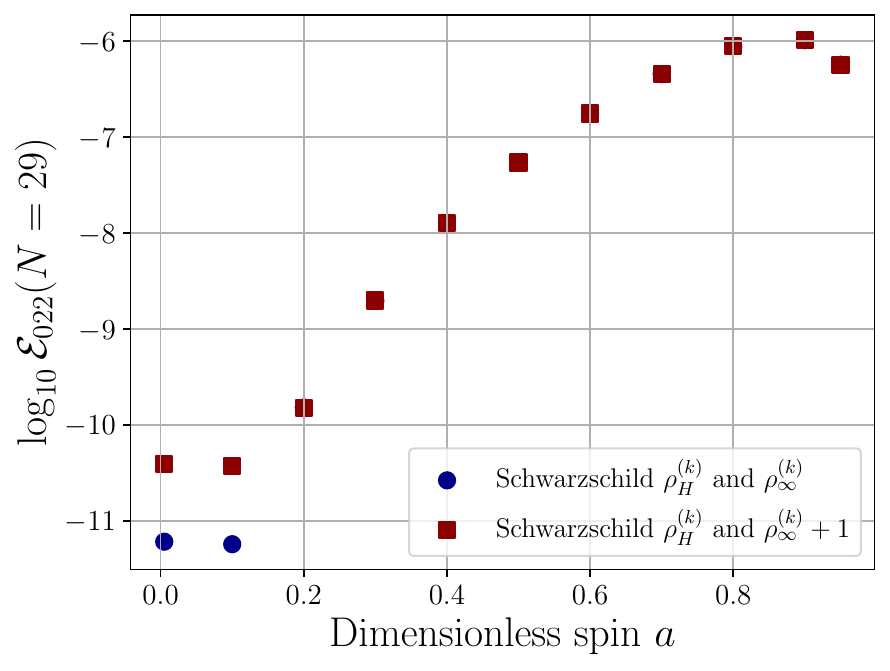}}
\caption{Absolute error of the 022-mode frequency using the METRICS approach with 30 Chebyshev and associated Legendre polynomials, with the Schwarzschild $\rho_H^{(k)}$ and $\rho_{\infty}^{(k)}$ (blue circles, see main text) and the same $\rho_H^{(k)}$ and $\rho_{\infty}^{(k)}$ but enlarged by unity (red squares). 
Observe that a different radial scaling does not significantly affect the accuracy of the 022-mode frequency, which indicates the robustness of the METRICS approaches against different scalings.
}
 \label{fig:Absolute_errors_rhos}
\end{figure}

As in the Schwarzschild case, we find that using different values of $\rho_H^{(k)}$ and $\rho_{\infty}^{(k)}$ still allows us to accurately compute the QNM frequencies of a Kerr BH, provided that $\rho_H^{(k)}$ and $\rho_{\infty}^{(k)}$ are large enough to capture the diverging behavior of the metric perturbations at the event horizon and spatial infinity. If this is the case, then the corresponding $u_k ( -1 \leq z, \chi \leq +1)$ are finite within the computational domain. 
Figure~\ref{fig:Absolute_errors_rhos} shows the absolute error of the METRICS 022-mode frequency computed using the Schwarzschild $\rho_H^{(k)}$ and $\rho_{\infty}^{(k)}$ (given by Eq.~\eqref{eq:rhos}, blue circles) and using the same $\rho_H^{(k)}$ and $\rho_{\infty}^{(k)}$ but enlarged by unity (red squares). 
Observe that the absolute error of different sets of $\rho_H^{(k)}$ and $\rho_{\infty}^{(k)}$ are very close to each other, indicating that the accuracy of the QNM frequency calculations does not sensitively depend on the explicit $\rho_H^{(k)}$ and $\rho_{\infty}^{(k)}$ used, which is observed and explained in \cite{Chung:2023zdq}.  
Moreover, as also pointed out in \cite{Chung:2023zdq}, this robustness allows us to circumvent uncertainty about the ``correct" $\rho_H^{(k)}$ and $\rho_{\infty}^{(k)}$, such as the dependence of $\rho_H^{(k)}$ and $\rho_{\infty}^{(k)}$ on the dimensionless spin (if any).

\begin{figure*}[htp!]
\centering  
\subfloat{\includegraphics[width=0.47\linewidth]{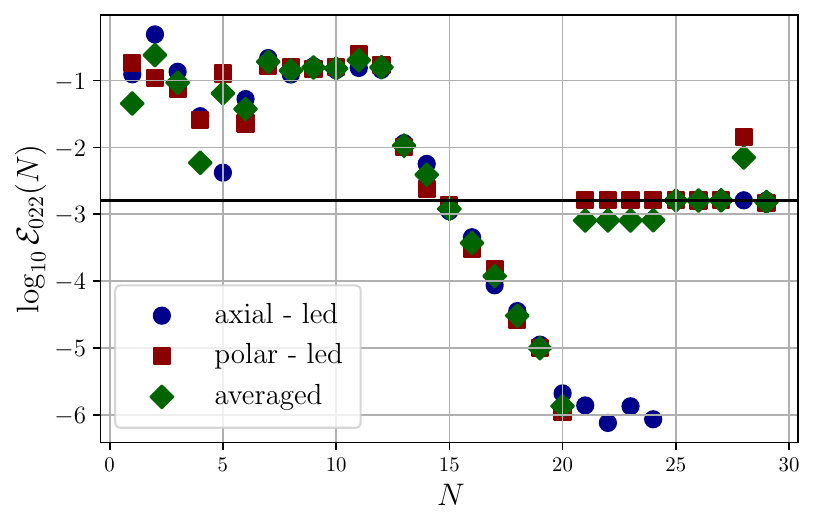}}
\subfloat{\includegraphics[width=0.47\linewidth]{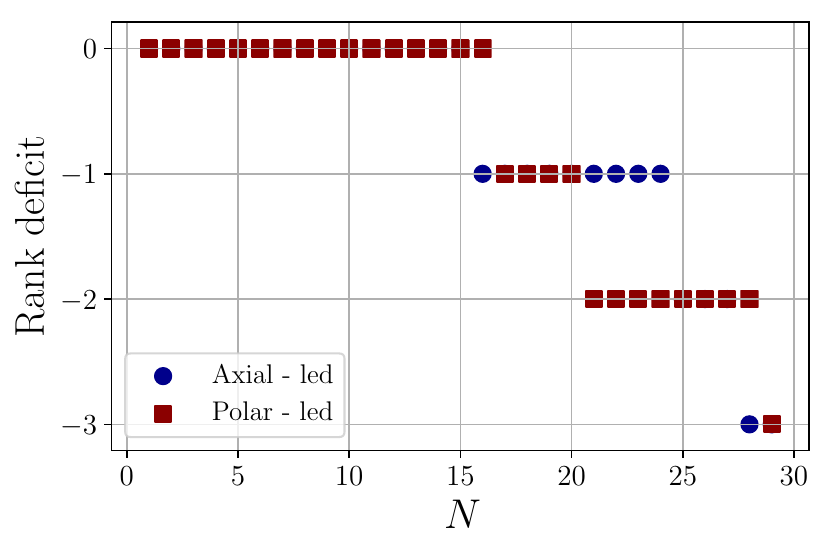}}
\caption{Absolute error of the axial- (blue circles), polar-led (red squares) and averaged (green diamonds) METRICS frequency of a Kerr BH with $a=0.9$, computed assuming $\rho_{H}^{(k)} = \rho_{\infty}^{(k)} = 2$, as a function of the spectral order $N$ (left panel), as well as the absolute error of the initial guess of the Newton-Raphson iterations (black horizontal line). 
Observe that the minimal absolute error is $\sim 10^{-6}$, which is similar to the minimal error computed using the other set of $\rho_{H}^{(k)} $ and $ \rho_{\infty}^{(k)} $. 
This indicates that the METRICS approach can still be used to accurately compute the QNM frequency of rapidly rotating BHs using component-independent $\rho_{H}^{(k)} $ and $ \rho_{\infty}^{(k)} $. 
However, observe also that when $N$ is large, the absolute error of the final results of the Newton-Raphson iterations is almost the same as the error of the initial guess, meaning that at that $N$ the Newton-Raphson iterations fail to improve our initial guess. 
We find that the Newton-Raphson iterations fail at the spectral order where the rank deficit (see Eq. \eqref{eq:rank_deficit} for definition) of the Jacobian matrix at the final step of the iterations decrease to $\leq -2 $ (right panel). 
This correlation suggests that we should also monitor the matrix rank as the iterations and to discard the results suffering from significant rank deficit. 
}
\label{fig:rank}
\end{figure*}

Other than these sets of $\rho_H^{(k)}$ and $\rho_{\infty}^{(k)}$, we also experiment with the component-independent set of $\rho_{H}^{(k)} = \rho_{\infty}^{(k)} = 2$. 
We find that this set of $\rho_H^{(k)}$ and $\rho_{\infty}^{(k)}$ still allows us to accurately compute the 022-mode frequency of the Kerr BH, but special care needs to be taken. 
We find that, for rapidly rotating BHs and large $N$, the Newton-Raphson iterator is trapped and stuck at the initial guess. 
In other words, the iterations do not significantly improve the initial guess. 
For example, the left panel of the Fig.~\ref{fig:rank} shows $\log_{10} \mathcal{E}_{022}(N)$ of the axial-, polar- and averaged 022-mode frequency as a function of $N$ computed with $\rho_{H}^{(k)} = \rho_{\infty}^{(k)} = 2$. 
The horizontal line marks the absolute error of the initial guess of the Newton-Rapshon iterations for computing the 022-mode frequency of a Kerr BH with $a=0.9$. 
We observe that for $N \lesssim 12$, $\mathcal{E}_{022}(N)$ first fluctuates between $\sim 10^{-2} - 10^{-1} $, and for $13 \leq N \leq 20 $, $\mathcal{E}_{022}(N)$ decreases exponentially. 
The absolute error at $N = 20$ is $\sim 10^{-6}$, which is similar to the accuracy achieved using the previous sets of $\rho_H^{(k)}$ and $\rho_{\infty}^{(k)}$. 
For $N > 20 $, $\mathcal{E}_{022}(N)$ of all three frequencies is very close to the error of the initial guess, which suggests that the iterations fail to improve our initial guess. 

This failure of the iteration method is related to the rank of the Jacobian matrix. 
Explicitly, when $N$ is sufficiently large, the rank of $\textbf{J}$ can be significantly smaller than the length of $\textbf{x}$. 
To illustrate this fact, the right panel of Fig.~\ref{fig:rank} shows the rank deficit of the Jacobian matrix, 
\begin{equation}\label{eq:rank_deficit}
\text{rank deficit} = \text{rank}(\textbf{J}) - \text{dim}({\textbf{x}}), 
\end{equation}
at the final step of the Newton-Raphson iterations as a function of the spectral order. 
We observe that when $N \geq 21 $ for the polar-led perturbations and $N \geq 25$ for the axial-led perturbations, which are the spectral order when the Newton-Raphson iterations of the corresponding leading parity fail, the rank deficit drops to $\leq -2 $, which implies that the system is significantly underdetermined. 
One possible explanation for why this does not happen with the $\rho_{H}^{(k)}$ and $\rho_{\infty}^{(k)}$ of the Schwarzschild spacetime, is that, by setting $\rho_{H}^{(k)}$ and $\rho_{\infty}^{(k)}$ to be the same number of all $k$, we are artificially demanding that perturbations of all metric components follow a similar asymptotic behavior, which increases the degeneracy of the problem. 
The correlation between the failure of the Newton-Raphson iterations and the rank deficit suggests that one should also monitor the rank of the Jacobian matrix to discard results obtained with Jacobian matrices of rank deficit $\leq -2$. 

\section{Extractions of metric perturbations with the METRICS approach}
\label{sec:metric_reconstructions}

The METRICS approach offers a distinct advantage over the established Teukolsky formalism when investigating perturbed GW metrics. 
In the Teukolsky formalism, obtaining metric perturbations from curvature perturbations involves a complex sequence of steps, including the introduction of intermediary quantities, such as the Hertz potential~\cite{Chrzanowski:1975wv,Kegeles_Cohen_1979,Lousto:2005xu}. 
On the other hand, the METRICS approach starts with metric perturbations, obviating the need for any supplementary metric reconstruction procedure.
In this section, we discuss the extraction of metric perturbations using the METRICS approach and validate the accuracy of the reconstructed metric (which we refer to as the ``spectral metric"). We shall do so by comparing the Teukolsky perturbation function ($\psi$, the solution of the Teukolsky equation) computed using the spectral metric to that obtained by numerically solving the Teukolsky equation. 
Finally, we analyze the relative content of the axial and polar sectors of metric perturbations of the Kerr BH with $a \leq 0.95$ obtained using the METRICS approach. 

\subsection{Numerical validation with the Teukolsky equations}
\label{sec:metric_numerical_validation}

We validate the accuracy of the spectral metric by comparing the Teukolsky perturbation function  computed using the spectral metric, $\psi (\text{spec})$, to that obtained using Leaver's method, $\psi (\rm L)$.
In this section, to illustrate the strength of the METRICS approach, we focus only on the metric perturbations of an $a=0.9$ Kerr BH, because it is sufficiently rapidly rotating to allow us to quantify any source of error in the metric perturbations. 

The Teukolsky perturbation function is related to the perturbed  
Weyl scalar $\psi_4$ via
\begin{equation}\label{eq:psi_psi_4_relation}
\psi = (r-i M a\cos \theta)^4 \psi_4, 
\end{equation}
and $\psi_4$ is related to metric perturbations through \cite{PhysRevD.11.2042, Whiting_Price_2005, Lousto:2005xu}
\begin{align}
\label{eq:psi_4_metric_relation}
 \psi_4 
& =  \frac{1}{2}\left\{(\hat{\delta}+3 \alpha+\bar{\beta}-\bar{\tau})(\hat{\delta}+2 \alpha+2 \bar{\beta}-\bar{\tau}) h_{\mathrm{nn}} \right. 
\nonumber \\
& \quad +(\hat{\triangle}+\bar{\mu}+3 \gamma-\bar{\gamma})(\hat{\triangle}+\bar{\mu}+2 \gamma-2 \bar{\gamma}) h_{\overline{\mathrm{mm}}}
\nonumber  \\
& \quad -[(\hat{\triangle}+\bar{\mu}+3 \gamma-\bar{\gamma})(\hat{\delta}-2 \bar{\tau}+2 \alpha) 
\nonumber \\
& \quad \left.+(\hat{\delta}+3 \alpha+\bar{\beta}-\bar{\tau})(\hat{\triangle}+2 \bar{\mu}+2 \gamma)] h_{(\mathrm{n} \overline{\mathrm{m}})} \right\},
\end{align}
where 
\begin{equation}
\begin{split}
 \hat{\delta} = \bar{m}^{\mu} \partial_{\mu}, \qquad
 \hat{\Delta} = n^{\mu} \partial_{\mu}, 
\end{split}
\end{equation}
$n^{\mu}$ and $\bar{m}^{\mu}$ are two of the Kinnersly null tetrads in the Boyer-Lindquist coordinates, the quantities $\alpha, \beta, \mu$, and $\tau$ are NP spin coefficients \cite{Teukolsky_01_PRL, Teukolsky:1973ha, Teukolsky:1974yv, Loutrel_Ripley_Giorgi_Pretorius_2020}, and
\begin{equation}
\begin{split}
h_{nn} & = h_{\mu \nu} n^{\mu} n^{\nu}, \\
h_{\overline{\mathrm{mm}}} & =h_{\mu \nu} \bar{m}^{\mu} \bar{m}^{\nu}, \\
h_{n \overline{\mathrm{m}}} & =h_{\mu \nu} n^{\mu} \bar{m}^{\nu},
\end{split}
\end{equation}
are the metric perturbations projected onto the tetrad vectors. 
We remind the reader that $\hat{\Delta}$ is different from $\Delta = (r-r_+)(r-r_-)$. 

To compute the Teukolsky perturbation function, $\psi(\text{spec})$ at a given spectral order $N$, we first compute the metric perturbation $h_{\mu \nu}$ using the METRICS approach (Eq.~\eqref{eq:spectral_decoposition_factorized_finite}). 
We then project $h_{\mu \nu}$ onto the null vectors $n^\mu$ and $\bar{m}^\mu$ to obtain the perturbed fourth Weyl scalar $\psi_4$ (Eq.~\eqref{eq:psi_4_metric_relation}). 
Finally, we compute $\psi(\text{spec})$ from $\psi_4$ using Eq.~\eqref{eq:psi_psi_4_relation}. We denote the $\psi(\text{spec})$ computed using the spectral metric up to $N$ spectral orders as $\psi(\text{spec}; N)$.

To obtain $\psi(\text{L})$, we solve the Teukolsky equation (with $s=-2$) using Leaver's, continued fraction method \cite{Leaver:1985ax}. More specifically, we solve the radial and angular Teukolsky equations using a 150-term continued fraction. 
We keep this number of terms because we find that the QNM frequency and the separation constant only change below the $10$th decimal place 
if more terms are included in the calculations (i.e. with a large $N$). 

\begin{figure*}[htp!]
\centering  
\subfloat{\includegraphics[width=0.47\linewidth]{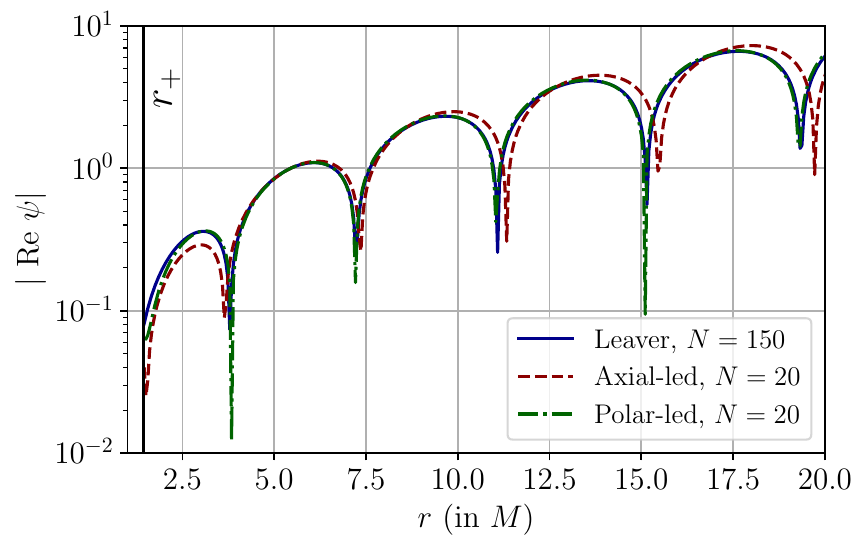}}
\subfloat{\includegraphics[width=0.47\linewidth]{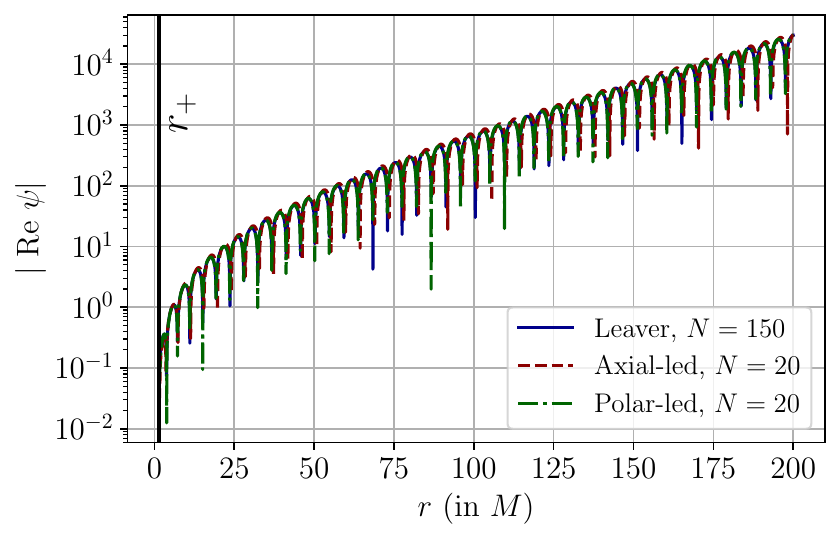}}
\qquad
\subfloat{\includegraphics[width=0.47\linewidth]{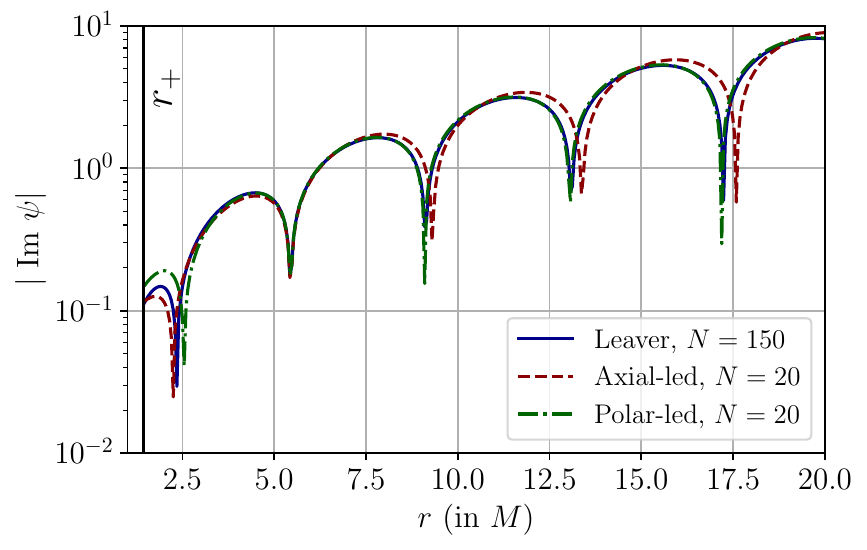}}
\subfloat{\includegraphics[width=0.47\linewidth]{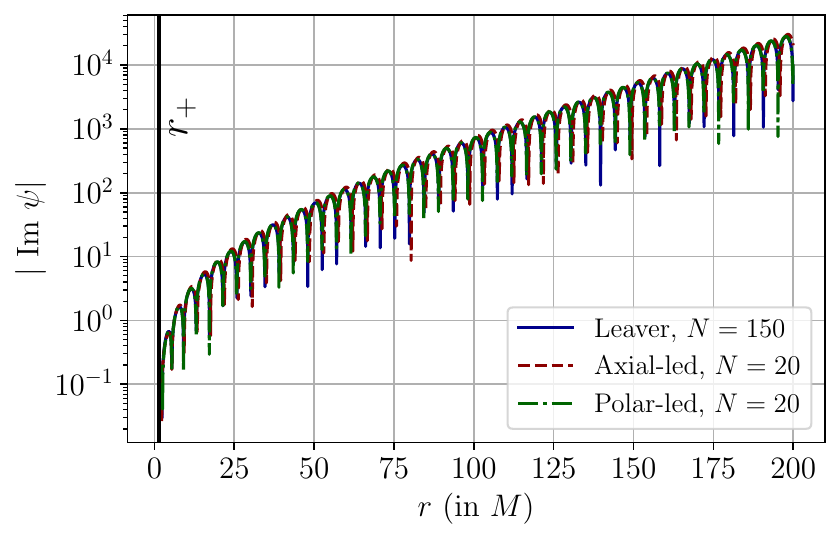}}
\caption{The absolute value of the real (top panels) and imaginary parts (bottom panel) of the Teukolsky perturbation $\psi = (r-iMa \cos \theta)^4 \psi_4$, where $\psi_4$ is the perturbed fourth-Weyl scalar, of a Kerr BH of dimensionless spin $a=0.9$, at $t=0$ and on the equatorial plane. 
The left panels cover from $r = r_+ + 10^{-2} M$ to $20 M$; the right panels cover from $r = r_+ + 10^{-2} M$ to $200 M$. 
In all panels, the vertical axis is in logarithmic scale, the solid vertical line in black marks the position of the outer event horizon ($r_+$), the solid blue line shows $\psi$ obtained by solving the Teukolsky equation using the Leaver continued fraction method, and the dashed red and dashed-dotted green lines show $\psi$ obtained using the axial- and polar-led metric perturbations reconstructed using the METRICS approach respectively.
Observe that the real and imaginary parts of all $\psi$ overlap almost completely with each other, show a similar variation pattern over $r$, have similar numerical values at a given $r$, and attain maxima and minima at similar $r$ positions. 
We evaluated that the mismatch (defined by Eq.~\eqref{eq:psi4_mm}) between the Leaver and the spectral $\psi_4$ to be $\sim 1 \%$ (see main text). 
These consistencies indicate that the METRICS approach can also accurately reconstruct metric perturbations around a rapidly rotating Kerr BH, while computing its QNM frequencies.
}
 \label{fig:Reconstructed_psi_4}
\end{figure*}

We then compare $\psi(\text{spec}; N)$ and $\psi(\text{L})$.
Since the Teukolsky equation is a homogenous, linear, partial differential equation, its solution admits linear-scaling invariance.
Thus, to make sure that we are comparing the $\psi(\text{spec}; N)$ and $\psi(\text{L})$ that correspond to the same Teukolsky function, without loss of generality, we fix,
\begin{equation}
\begin{split}
& \psi \left(\text{spec}; N, x^{\mu}_0 \right) = \psi \left(\text{L}, x^{\mu}_0 \right) = 1, \\
& x^{\mu}_0 = (t, r, \theta, \phi) = \left(0, 3, \frac{\pi}{2}, 0 \right),  
\end{split}
\end{equation}
by normalizing each of them by their value at $x_{\mu}^{0}$.
If the METRICS approach correctly reconstructs the metric perturbations around a Kerr BH, then $\psi(\text{spec})$ and $\psi(\text{L})$ should be the same.
Figure~\ref{fig:Reconstructed_psi_4} shows the real (right panel) and imaginary parts (left panel) of $\psi(\text{L})$ (solid blue line) and $\psi(\text{spec})$ (dashed red and dashed-dotted lines for the axial- and polar-led perturbations respectively).
Both $\psi(\text{spec})$ and $\psi(\text{L})$ vary with $r$ in a similar way. 
They attain maxima and minima at similar $r$ positions, and the numerical value of $\psi(\text{spec})$ at a given $r$ coordinate is close to that of $\psi(\text{L})$ at the corresponding radial coordinate. 
These consistencies indicate that the METRICS approach can correctly reconstruct the metric perturbations around the Kerr BH.

To quantify the differences between $\psi(\rm spec)$ and $\psi(\rm L)$ after the scaling, we define the following mismatch with respect to $r$, 
\begin{equation}\label{eq:psi4_mm}
\mathcal{MM} = 1 - \frac{\text{Re}[\psi(\text{spec})|\psi(\text{L})]}{[\psi(\text{spec})|\psi(\text{spec})]^{1/2} [\psi(\text{L})|\psi(\text{L})]^{1/2}}, 
\end{equation}
where 
\begin{equation}
\begin{split}
[A|B] = \int_{r=r_{\rm inf}}^{r=r_{\rm sup}} dr & A\left(t = 0, r, \theta = \frac{\pi}{2}, \phi \right) \\
& \times B^{*} \left(t = 0, r, \theta = \frac{\pi}{2}, \phi \right). 
\end{split}
\end{equation}
Here $r_{\rm inf} $ and $r_{\rm sup} $ should formally be $r_+$ and $\infty$, respectively, and the asterisk stands for complex conjugation.
For actual numerical evaluation, we slightly displace the lower limit from the event horizon to $r_{\rm inf} = r_+ + 10^{-2} $ to avoid overflow due to the divergence of the asymptotic factor multiplied to the metric; similarly, we set $r_{\rm sup} = r_{\rm inf} + 200 M $ to make sure the far-field behavior of $\psi$ is included in the calculations. 
To reduce the computational time required for this calculation, we focus on evaluating the 2 norm on the equatorial plane;
we expect the $\mathcal{M}$ evaluated at different $\theta$ will show similar variation with respect to $N$. 
If $\psi(\text{spec}) = \psi(\text{L})$, then $\mathcal{MM} = 0 $; if $\psi(\text{spec})$ is very different from $\psi(\text{L})$, then $\mathcal{MM} \sim 1 $. 
Thus, an $\mathcal{MM} $ closer to 0 indicates that $\psi(\rm spec)$ is similar to $\psi(L)$. 
We numerically evaluate the integrals as a Riemann sum at a radial resolution of $\Delta r = \frac{M}{2}$. 
and check that further reducing the resolution does not change the first three digits of the base-10 logarithms of $\mathcal{MM}$. 
We find that $\mathcal{MM} \sim 10^{-2}$ for both the axial- and polar-led perturbations of a Kerr BH with $a=0.9$, reconstructed using the METRICS approach. This implies that, in spite of the small differences, the mismatch between the Teukolsky function is very good.  

Finally, we conclude this subsection by comparing the asymptotic behavior of $\psi(\rm spec)$ and $\psi(\rm L)$ via asymptotic expansions. 
The asymptotic behavior of $\psi(\rm L)$ can be obtained by the asymptotic expansion of Leaver's, continued fraction solution to the Teukolsky equation. The asymptotic behavior of $\psi(\rm spec)$ can be obtained by considering the asymptotic expansion of $\psi$ that follows from our metric ansatz (Eq.~\eqref{eq:metpert}).
We find that at spatial infinity, both $\psi(\rm spec)$ and $\psi(\rm L)$ are asymptotic to $r^{3+2i\omega} e^{i \omega r}$, and thus, the asymptotic behaviors match exactly.  
At the event horizon, however, we find that 
\begin{equation}
\psi(\text{spec}) \sim \left( \frac{r-r_+}{r}\right)^{-2-i M (\omega - m \Omega_{H}) \frac{1+b}{b}}, 
\end{equation}
while
\begin{equation}
\psi(\text{L}) \rightarrow \left( \frac{r-r_+}{r-r_-}\right)^{-2-i M (\omega - m \Omega_{H}) \frac{1+b}{b}}. 
\end{equation}
The asymptotic behavior of $\psi(\rm spec)$ and $\psi(\rm L)$ at the event horizon differs by a factor of 
\begin{equation}
\left( \frac{r}{r-r_-}\right)^{-2-i M (\omega - m \Omega_{H}) \frac{1+b}{b}}, 
\end{equation}
which is finite at the event horizon.
Hence, $\psi(\rm spec)$ and $\psi(\rm L)$ have the same asymptotic behavior at the event horizon, up to a constant.
These comparisons analytically prove the validity of our metric ansatz (Eq.~\eqref{eq:metpert}) and our metric reconstruction procedures. 

\subsection{Parity content of the reconstructed metric perturbations}

\begin{figure}[tp!]
\centering  
\subfloat{\includegraphics[width=\columnwidth]{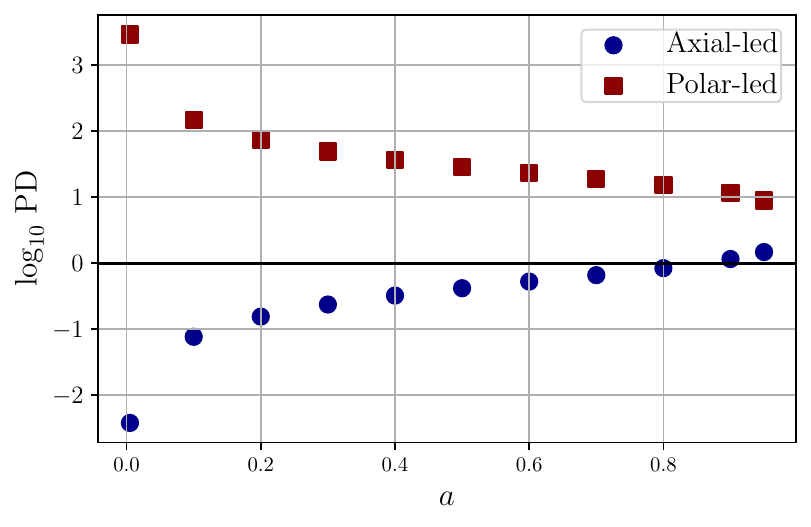}}
\caption{The base-10 logarithms of the parity dominance (defined by Eq.~\eqref{eq:PD_01}) of the reconstructed axial-led (blue circles) and polar-led (red squares) metric perturbations. 
The parity dominance, a non-negative number, gives a quantitative estimate of the ratio between the amplitude of the axial- and polar-led perturbation.
A PD $\gg 1$ indicates that the perturbations are mostly led by the polar sector, and a PD $\ll 1$ indicates that the perturbations are mostly led by the axial sector. 
Observe that for $a\lessapprox 0.5$, PD $\ll 1 $ for the axial-led perturbations and PD $\gg 1 $ for the polar-led perturbations, indicating that the axial-led perturbations indeed consist of mostly axial perturbations and similarly for the polar-led perturbations. 
}
 \label{fig:PD}
\end{figure}

The spectral method can also reconstruct metric perturbations that are mainly led by a particular parity. 
To quantify the parity content, we define the parity dominance (PD) for the spectral metric, 
\begin{equation}\label{eq:PD_01}
\begin{split}
\text{PD} & = \left( \frac{\displaystyle \sum_{k=1}^{4} \| u_k(z, \chi. N)\|_{W2}}{\displaystyle \sum_{k=5, 6} \| u_k(z, \chi, N)\|_{W2}}\right)^{\frac{1}{2}}, 
\end{split}
\end{equation} 
where $N$ is the spectral order of computing the QNM frequencies, and $\| . \|_{W2}$ is the weighted 2 norm, defined as 
\begin{equation}\label{eq:W2_norm}
\begin{split}
& \| f\|^2_{W2} = \int_{-1}^{+1} d\chi \int_{-1}^{+1} dz |f(z, \chi)|^2\left( 1 - z^2 \right)^{-\frac{1}{2}}. 
\end{split}
\end{equation}
Note that the weighted 2 norm differs from the usual 2 norm of a function by the inclusion of the weighted function $(1-z^2)^{-\frac{1}{2}}$, so that we can analytically evaluate the norm using the orthogonal properties of the Chebyshev and associated Legendre polynomials, 
\begin{align}
\label{eq:PD_02}
\text{PD} 
& = \left(\frac{\displaystyle \sum_{k=1}^{4} \sum_{n=0}^{N} \sum_{\ell = |m|}^{N+|m|} \frac{(\ell+m)!}{(2\ell+1)(\ell-m)!} |v_k^{n \ell}(N)|^2}{\displaystyle \sum_{k=5, 6} \sum_{n=0}^{N} \sum_{\ell = |m|}^{N+|m|} \frac{(\ell+m)!}{(2\ell+1)(\ell-m)!} |v_k^{n \ell}(N)|^2}\right)^{\frac{1}{2}}. 
\end{align}
Heuristically, the parity dominance provides an estimate of the ratio between the amplitude of the axial- and polar-led perturbations. 
If the reconstructed perturbations are purely axial, then $\text{PD}^{\rm (A)} = 0$; If the reconstructed perturbations are purely polar, then $\text{PD}^{\rm (A)} \rightarrow + \infty$.  

Figure~\ref{fig:PD} shows the base-10 logarithms of the PD of the spectral metric perturbations as a function of $a$. 
The horizontal line marks the line $\rm PD = 1 \Rightarrow \log_{10} PD = 0 $, when the amplitude of the axial- and polar-led perturbations is approximately equal. 
Observe that for $a \lessapprox 0.9$, PD $ < 0.1 $ for the axial-led perturbations, indicating that the axial-led perturbations indeed consist of mostly the axial perturbations; PD $ > 10 $ for the polar-led perturbations, indicating that the polar-led perturbations indeed consist of mostly the polar perturbations. 
Although the metric perturbations obtained here are coupled, one can show that, at all orders in spin, the even and odd parity versions of the Teukolsky master function decouple completely \cite{Li:Preparation}.
This decoupling implies that there may exist some initial guesses of the eigenvector $\textbf{x}$ to construct purely axial and polar perturbations of a rotating BH, which will be important for the computation of the QNM frequencies of BHs beyond GR, where isospectrality is typically broken \cite{Wagle:2021tam, QNM_dCS_02, QNM_dCS_03, QNM_dCS_04, QNM_EdGB_01, QNM_EdGB_02, QNM_EdGB_03}. 
Such initial guesses might be obtained by studying the transformation properties of the metric perturbation ansatz upon the action of the parity operator. 
We will explore such initial guesses in future work. 
Nonetheless, being able to compute metric perturbations that are led by a given parity when $a \lessapprox 0.9$ is still insightful and highly nontrivial.

\section{Concluding remarks}
\label{sec:conclusion}

We have here further developed and extended a novel spectral method for computing the frequencies of gravitational QNMs of rotating BHs. We have verified that our method can accurately and efficiently compute the frequency of the fundamental corotating quadrupole mode (the "022" mode) of a Kerr BH with dimensionless spin $a \leq 0.95$. 
The individual relative fractional errors of the real and imaginary parts of the frequency are less than $10^{-5}$. 
This accuracy is sufficient to analyze GW ringdown LIGO-Virgo data, because it is better by three orders of magnitude than the relative fractional uncertainty of the measured real and imaginary parts of the 022-mode frequency obtained by combining all ringdown signals \cite{LIGOScientific:2021sio}.
This accuracy should also be sufficient to analyze the black-hole ringdown signals detected by the next-generation detectors, whose relative fractional uncertainty of the measured real and imaginary parts of the 022-mode frequency is predicted to be $\leq 10^{-4}$ \cite{Maselli:2023khq}.
We have also demonstrated that our method calculates directly and accurately the metric perturbations along with the QNM frequencies and we have carried out a plethora of checks to verify the robustness of the method. 

The METRICS approach has several major advantages over other existing methods for computing the gravitational QNM frequencies of BHs.
First, the spectral method does not require the decoupling or simplification of the linearized field equations into several master equations. 
This is a significant advantage because decoupling may not be possible for perturbations of BHs surrounded by matter, scalar or vector fields and for BHs outside of GR.
Second, the METRICS approach is based on the Regge-Wheeler gauge, a common gauge in BH perturbation theory that has the advantage of greatly simplifying the linearized field equations.
This gauge is expected to exist in a wide class of modified gravity theories, such as dynamical Chern-Simons gravity \cite{QNM_dCS_01, QNM_dCS_02, QNM_dCS_03, QNM_dCS_04} and Einstein-dilaton Gauss-Bonnet gravity \cite{QNM_EdGB_01, QNM_EdGB_02, QNM_EdGB_03}, making the METRICS approach more broadly applicable than other methods based on different gauges (e.g., \cite{Dolan:2021ijg, Ripley:2020xby, Loutrel_Ripley_Giorgi_Pretorius_2020, Spectral_04, Spectral_05, Spectral_06, Dias:2014eua, Dias:2013sdc, Santos:2015iua, Dias:2015nua}).
Third, our spectral method can simultaneously reconstruct metric perturbations, while computing the QNM frequencies accurately. 
This implies the METRICS has direct access to the GW metric perturbation, without needing to integrate twice to convert between the Newman-Penrose scalars and the GW observable. 
These advantages make the METRICS approach a powerful tool for studying BH perturbations in and outside GR. 

The METRICS approach also provides new insights into the gravitational perturbations of the Kerr BH.
First, our work is the first explicit demonstration that the Regge-Wheeler gauge can be used to accurately compute the QNM frequencies of a rapidly rotating BH. 
Before this work, the Regge-Wheeler gauge had mostly been applied to Schwarzschild or slowly rotating BHs in GR or alternative theories \cite{QNM_dCS_01, QNM_EdGB_02, QNM_EdGB_03}. 
For rapidly-rotating BHs, previous studies relied heavily on the ingoing or outgoing radiation gauge through the application of the Teukolsky equation, or on the harmonic gauge for studies of the stability and thermal properties of BHs (e.g. \cite{Santos:2015iua, Monteiro:2009ke, Dias:2009iu, Dias:2010eu, Dias:2013sdc, Dias:2014eua, Dias:2015nua, Dias:2015wqa, Dias:2018ufh, Dias:2021yju, Dias:2022oqm, Dias:2022str}). 
Compared to other gauges, the Regge-Wheeler gauge is particularly convenient because it can be easily enforced in the frequency domain by setting some perturbation components to zero. 
This reduces the complexity and length of the linearized field equations, especially if one works directly with metric perturbations instead of curvature perturbations. 

A second insight into gravitational perturbations of Kerr BHs that is derived from our work is related to our use of associated Legendre polynomials. We show that this class of elementary and commonly used spectral functions are sufficient for studying the gravitational perturbations of rotating BHs. 
Usually, associated Legendre polynomials are used to study scalar perturbations of rotating BHs or the gravitational perturbations of slowly rotating BHs. 
Inspired by the Teukolsky equation, a more ``natural" choice of the angular spectral basis would be spheroidal harmonics or spin-weighted spherical 
harmonics, but these bases are more mathematically complicated. 
Using the associated Legendre polynomials can reduce the difficulty of solving the linearized field equations because this class of spectral functions is simpler, more elementary, and more familiar to physicists.

Further refinements of the spectral method can be explored in future applications to improve its accuracy and speed. 
One possible refinement is to use a more sophisticated spectral basis, such as spin-weighted spherical
harmonics for the angular coordinate. 
Another possible refinement is to implement a more sophisticated variant of the Newton-Raphson iterative scheme to improve its performance. 
Yet another refinement is to use of resummation techniques on the metric perturbations to improve the accuracy of the QNM frequency calculations see, e.g., \cite{QNM_EdGB_03}. 
The results reported in the paper, however, show that the METRICS approach, in its current incarnation, is already sufficiently accurate and robust to produce results that are applicable to the analyses of actual ringdown signals.

Our immediate next step is to apply the spectral method to study the gravitational QNMs of rapidly rotating BH in modified gravity theories, but several other applications are possible. 
One such application is to study the perturbations of spinning BHs in the presence of matter, in order to use ringdown signals as a probe of dark matter with future space-based detectors. 
Another application is the generalization of the METRICS approach to model BHs in the presence of external sources, as in the case of extreme mass-ratio inspirals. This could be done by working in the frequency domain and generalizing the algebraic matrix equation to an inhomogeneous one. 
These applications and more render the METRICS approach a powerful tool for studying the perturbations of BHs with any spin both in and outside general relativity, both in vacuum or in the presence of external matter perturbations. 

\section*{Acknowledgements}

The authors acknowledge the support from the Simons Foundation through Award No. 896696, the NSF through Award No. PHY-2207650 and NASA through Grant No. 80NSSC22K0806. 
The authors would like to thank Emanuele Berti, Yanbei Chen, Mark H.Y. Cheung, Dongjun Li, Tjonnie Li, Justin Ripley and Leo Stein for insightful discussions, and Gregorio Carullo, Kwinten Fransen, Simon Maenaut and Shinji Tsujikawa for comments on the initial version of the manuscript.
A.K.W.C would like to thank Alan Tsz Lok Lam and Lap Ming Lin for useful advice offered at the beginning of this work, and Nijaid Arredondo for catching a typo. 
N.Y. would like to thank Takahiro Tanaka for insightful discussion about this work. 
The calculations and results reported in this paper were produced using the computational resources of the Illinois Campus Cluster, a computing resource that is operated by the Illinois Campus Cluster Program (ICCP) in conjunction with NCSA, and is supported by funds from the University of Illinois at Urbana-Champaign. 
The author would like to specially thank the investors of the GravityTheory, CAPs, astrophysics, and physics computational nodes for permitting the authors to execute runs related to this work using the relevant computational resources. 

\appendix

\section{Symbols}
\label{sec:Appendix_A}

The calculations presented in this paper involved numerous symbols. 
For convenience of the reader, we provide a list of the symbols and their definitions in this appendix. 

\begin{itemize}
    \item $a$ is the dimensionless spin of the BH, first defined in Eq.~\eqref{eq:metric}. 
    \item $ A_k (r) $ is the asymptotic prefactor of the $k$-th perturbation variable, first defined in Eq.~\eqref{eq:asym_prefactor}. 
    \item (A) is the superscript which denotes the quantity concerning the axial-led perturbations, first defined in Eq.~\eqref{eq:vector_equations_w_axial_convention}. 
    \item $b = \sqrt{1-a^2}$, first defined in Eq.~\eqref{eq:metric_quantities}. 
    \item $\mathcal{B}(N)$ is the backward modulus difference of the QNM frequency, first defined in Eq.~\eqref{eq:backward_modulus_diff}. 
    \item $d_{r}$ is the degree of $r$ of the coefficient of the partial derivative of the linearized Einstein equations, first defined in Eq.~\eqref{eq:pertFE-1}. 
    \item $d_{\chi}$ is the degree of $\chi$ of the coefficient of the partial derivative of the linearized Einstein equations, first defined in Eq.~\eqref{eq:pertFE-1}. 
    \item $d_{z}$ is the degree of $z$ of the coefficient of the partial derivative of the compactified linearized Einstein equations, first defined in Eq.~\eqref{eq:system_3}. 
    \item $\mathbb{D}(\omega) $ is the coefficient matrix of spectral expansion, from one particular basis to another, first defined in Eq.~\eqref{eq:pertFE-2}. 
    \item $\Delta = (r-r_+)(r-r_-)$, first defined in Eq.~\eqref{eq:metric_quantities}. 
    \item $\Delta^{\rm Re/Im}$ is the relative fractional error in the real and imaginary parts of the METRICS QNM frequencies, $\omega(\text{METRICS})$, and the Teukolsky equations, first defined in Eq.~\eqref{eq:relative_error}. 
    \item $\mathcal{E}(N)$ is the absolute error between the METRICS QNM frequencies, $\omega(\text{METRICS})$, and Leaver's method to solve for the QNM modes $\omega (\text{L})$, first defined in Eq.~\eqref{eq:Error_rho_1}. 
    \item $f_k$ is the $k$-th algebraic equation which we solve to reconstruct metric perturbations, first defined in Eq.~\eqref{eq:vector_equations_w_polar_convention}. 
    \item $\mathcal{G}_{k, \gamma, \delta, \sigma, \alpha, \beta, j} $ is the coefficient of $\omega^\gamma r^{\delta} \chi^{\sigma} \partial_{r}^{\alpha} \partial_{\chi}^{\beta} h_j$ of the linearized Einstein equations of $h_j$, first defined in Eqs.~\eqref{eq:pertFE-1}. 
    \item $h_k(r, \chi)$ is the functions of metric perturbations, first defined in Eqs.~\eqref{eq:odd} and \eqref{eq:even}. 
    \item $i = \sqrt{-1}$ is the imaginary unit. 
    \item $k$ in the subscript is the component of the metric perturbation functions and $k= 1, ..., 6$, first defined in Eqs.~\eqref{eq:odd} and \eqref{eq:even}.
    \item $\mathcal{K}_{k, \alpha, \beta, \gamma, \delta, \sigma, j} $ is the coefficient of $\omega^\gamma z^{\delta} \chi^{\sigma} \partial_{z}^{\alpha} \partial_{\chi}^{\beta}(...)$ of the linearized Einstein equations in $z$ and $\chi$, first defined in Eq.~\eqref{eq:system_3}. 
    \item $l$ is the azimuthal mode number of the gravitational QNMs, first defined in Sec.~\ref{sec:intro}. 
    \item $\ell$ is the degree of associate Legendre polynomial used in spectral expansion, first defined in Eq.~\eqref{eq:spectral_decoposition_correction_factor}.
    \item $M$ is the BH mass, which is taken to be $M=1$ throughout this work, first defined in Eq.~\eqref{eq:metric}. 
    \item $m$ is the azimuthal number of the metric perturbations, first defined in Eqs.~\eqref{eq:odd} and \eqref{eq:even}. 
    \item $N$ is the number of the Chebyshev and associated Legendre polynomials used in the full spectral expansion, first defined in Sec.~\ref{sec:setup}. 
    \item $\mathcal{N}_{\chi}$ is the number of the associated Legendre polynomials included in the spectral expansion, first defined in Eq.~\eqref{eq:spectral_decoposition_factorized_finite}. 
    \item $\mathcal{N}_{z}$ is the number of the Chebyshev polynomials included in the spectral expansion, first defined in Eq.~\eqref{eq:spectral_decoposition_factorized_finite}. 
    \item (P) is the superscript which denotes the quantity concerning the parity-led perturbations, first defined in Eq.~\eqref{eq:vector_equations_w_polar_convention}.
    \item PD is the parity dominance, which characterizes the parity content of metric perturbations, first defined in Eq.~\eqref{eq:PD_01}.
    \item $r_\pm = M(1\pm\sqrt{1-a^2})$ is the radial coordinate of the position of the event horizon of the Kerr BH, first defined below Eq.~\eqref{eq:metric_quantities}. 
    \item $r_*$ is the tortoise coordinate, first defined in Eq.~\eqref{eq:r_*}. 
    \item $\rho_{\infty}^{(k)}$ and $\rho_H^{(k)}$ are the parameters that characterize the boundary conditions of $h_k$ in future null infinity and at the horizon, first defined in Eq.~\eqref{eq:asymptotic_limits1} and \eqref{eq:asymptotic_limits2}. 
    \item $\omega_{q}(\rm L)$ is the frequency of the QNM $q$ computed using the Leaver method, first defined in Eq.~\eqref{eq:Error_rho_1}. 
    \item $z = \frac{2 r_+}{r} - 1$ is the variable that maps $r$ into a finite domain, first defined in Eq.~\eqref{eq:z}. 
\end{itemize}

\bibliography{ref}

\end{document}